\documentclass[11pt,a4paper]{article}

\newif\ifjhep
\IfFileExists{jheppub.sty}{\jheptrue}{\jhepfalse}

\ifjhep
  \usepackage{jheppub}
\else
  \usepackage[margin=2.6cm]{geometry}
  \newcommand{\affiliation}[1]{\gdef\savedaffiliation{#1}}
  \newcommand{\emailAdd}[1]{\gdef\savedemail{#1}}
  \newcommand{\keywords}[1]{}
  \renewcommand{\abstract}[1]{\gdef\savedabstract{#1}}
\fi

\usepackage{amsmath,amssymb,amsthm}
\ifjhep\else\usepackage{hyperref}\fi
\usepackage{tikz}
\usepackage[capitalise]{cleveref}

\newif\ifdraftlabels
\draftlabelsfalse
\ifdraftlabels\usepackage[notref,notcite,color]{showkeys}\fi

\theoremstyle{plain}
\newtheorem{theorem}{Theorem}[section]
\newtheorem{lemma}[theorem]{Lemma}
\newtheorem{corollary}[theorem]{Corollary}
\crefname{theorem}{Theorem}{Theorems}
\Crefname{theorem}{Theorem}{Theorems}
\crefname{lemma}{Lemma}{Lemmas}
\Crefname{lemma}{Lemma}{Lemmas}

\newcommand{\ads}{\mathrm{AdS}}

\newcommand{\cA}{\mathcal{A}}
\newcommand{\cB}{\mathcal{B}}
\newcommand{\cC}{\mathcal{C}}
\newcommand{\cG}{\mathcal{G}}

\DeclareMathOperator{\Arg}{Arg}

\title{\boldmath Only the disk: topological obstructions for warped $\ads_6\times S^2$ in Type IIB}

\author{Nakwoo Kim}
\affiliation{Department of Physics and Research Institute of Basic Science,\\
Kyung Hee University, Seoul 02447, Republic of Korea}
\emailAdd{nkim@khu.ac.kr}

\abstract{%
Supersymmetric $\ads_6\times S^2$ solutions of Type IIB supergravity are warped
products over a Riemann surface $\Sigma$. They are determined by two holomorphic
differentials on $\Sigma$, with five-branes at poles on its boundary.
We derive restrictions on the genus and the number of boundary components of $\Sigma$.
Without seven-branes or orientifolds, the differentials are
taken to be single-valued on $\Sigma$. Regularity implies that the number of interior zeros of one differential equals
the total rotation of the phase of the ratio of the two differentials along the boundary.
It also bounds this rotation on each boundary component by the number of poles on that component.
Combined with the known count of zeros on the double of $\Sigma$, these two statements
force $\Sigma$ to be a disk. With mutually local seven-branes and orientifolds as
interior punctures, the same conclusion follows under additional assumptions on
the global structure of the differentials and on their behavior at the punctures.
The disk then admits at most one orientifold puncture.
On a disk without punctures, the bound applied between successive poles
also recovers the known cyclic ordering of the five-brane charges along the boundary.
}


\begin{document}
\maketitle

\ifjhep\else
\begin{center}\small\emph{\savedaffiliation}\\[2pt]\texttt{\savedemail}\end{center}
\vspace{0.5em}
\begin{center}\begin{minipage}{0.92\textwidth}\small\noindent
\textbf{Abstract.} \savedabstract
\end{minipage}\end{center}
\vspace{1.5em}
\fi

\section{Introduction}
\label{sec:intro}

Type IIB supergravity admits solutions with sixteen supercharges and
$SO(2,5)\times SO(3)$ isometry. These provide holographic duals for five-dimensional
superconformal field theories. A large class of these theories is realized by intersections of $(p,q)$
five-branes, with or without additional seven-branes
\cite{Aharony:1997ju,Aharony:1997bh,DeWolfe:1999hj}. The local supersymmetry conditions
and the $S^2$ fibration over a two-dimensional base were derived using pure spinors by
Apruzzi et al.\ \cite{Apruzzi:2014qva} and recovered through a Killing spinor analysis by
Kim, Kim and Suh \cite{Kim:2015AdS6}. We refer to the three papers in the series
\emph{Warped $\ads_6\times S^2$ in Type IIB supergravity} as DGKU~I,
DGU~II and DGU~III. In DGKU~I, D'Hoker, Gutperle, Karch and Uhlemann
obtained the general local solution in terms of holomorphic data
\cite{DHoker:2016ujz}.\footnote{For an extension to complexified Type IIB
supergravity and a classification of its reality conditions, see
\cite{DHoker:2025nid}. The analytic continuation from $\ads_6\times S^2$ to $S^6\times S^2$
gives the Euclidean solutions dual to the polarized IKKT matrix model
\cite[App.~A]{Komatsu:2024bop}.} D'Hoker, Gutperle and Uhlemann constructed global
solutions for five-brane webs \cite{DHoker:2016ysh}, with the full construction in DGU~II
\cite{DHoker:2017mds}, and incorporated mutually local seven-branes in DGU~III
\cite{DHoker:2017zwj}.
In these solutions the metric is
\begin{equation}
  ds^2=f_6^2\,ds^2_{\ads_6}+f_2^2\,ds^2_{S^2}+4\rho^2|dw|^2 ,
  \label{eq:ansatz}
\end{equation}
where $w$ is a local coordinate on a Riemann surface $\Sigma$ with boundary, and $f_6$, $f_2$
and $\rho$ are real functions on $\Sigma$. The solutions
realize the exceptional superalgebra $F(4)$ and are specified by two locally holomorphic
functions $\cA_\pm(w)$. Poles of their differentials $d\cA_\pm=\partial_w\cA_\pm\,dw$ on $\partial\Sigma$ correspond to external
five-branes, with charges determined by the residues. Interior punctures carry the monodromies
of seven-branes.

In the global solutions constructed in these works, $\Sigma$ is a disk.
These disk solutions have enabled computations of free energies and entanglement entropies
\cite{Gutperle:2017tjo}, operator spectra \cite{Gutperle:2018wuk,Bergman:2018hin},
Wilson loops \cite{Uhlemann:2020bek} and the Euclidean on-shell action \cite{Adhikari:2026xdb}.
The holographic sphere free energies and stress-tensor coefficients 
$C_T$ are found to agree at large charges with localization results for the $T_N$ 
and $+_{N,M}$ theories \cite{Fluder:PrecisionTest}.
Disk solutions have also been used to give evidence for a five-dimensional $F$-theorem \cite{Fluder:2020pym}
and to study long quivers \cite{Uhlemann:2019ypp} and their mass deformations
\cite{Akhond:2022awd,Akhond:2022oaf}. Further developments incorporate orientifold
planes \cite{Uhlemann:2019lge} and holographic defects \cite{Gutperle:2020rty,Santilli:2023fuh},
while Penrose limits probe the geometry near null geodesics \cite{Gutperle:2021nkl}.

DGU~II formulated the regularity conditions for arbitrary genus and any
nonzero number of boundary components \cite[\S6]{DHoker:2017mds}.
Their numerical search on the unpunctured annulus suggested an obstruction
\cite[\S5.7]{DHoker:2017mds}, leaving open which topologies admit regular solutions.
The question also bears on the comparison with four-dimensional class $\mathcal S$
theories. Their holographic duals involve M5-branes wrapped on a Riemann surface whose
genus and complex structure can be varied \cite{Fenati:2026hkn}. We return
to this comparison in \cref{sec:discussion}.

For the known disk solutions, Kaidi and Uhlemann identified the Schottky
double $\hat\Sigma$, obtained by gluing $\Sigma$ to its mirror along the
boundary, with the M-theory curve of the corresponding five-brane junction \cite{Kaidi:2018zkx}. They checked
this correspondence for several explicit families.
If the identification extends to other topologies, a handle or an additional
boundary component of $\Sigma$ would give an M-theory curve of positive genus.
Kaidi and Uhlemann associated positive-genus curves with five-brane webs
whose internal faces have finite size. They interpreted these configurations
as mass deformations and hence expected only the disk to occur at the
conformal fixed point. A finite face can also arise at fixed masses from a
Coulomb branch expectation value, which likewise introduces a scale
\cite{Aharony:1997bh}.

Our main result establishes the restriction to the disk directly from Type IIB regularity.
In the monodromy-free class of DGU~II, let $\Sigma$ have genus $g$,
$\nu$ boundary components and $L$ boundary five-brane poles, with no interior
punctures and single-valued meromorphic differentials $d\cA_\pm$. We prove that
physical regularity, in the sense of \cref{sec:global}, requires
\begin{equation}
  2g+\nu<2,
  \qquad\text{that is}\qquad
  (g,\nu)=(0,1):\ \Sigma\ \text{is a disk} .
  \label{eq:main}
\end{equation}
In particular, the annulus is excluded for arbitrary charges and distributions
of the poles between its two boundaries.

For surfaces with mutually local seven-brane punctures, each carrying a parabolic
monodromy, \cref{thm:sevenbranes} again gives $(g,\nu)=(0,1)$.
Mutual locality restricts the seven-brane charges to a common direction.
Supersymmetric brane webs also admit mutually non-local seven-branes
\cite{DeWolfe:1999hj}, which we do not consider in this paper.
We assume that the DGU~III construction holds globally, with no independent
duality holonomy, that is, no duality transformation beyond those generated by
the punctures, and that the monodromy-invariant combination of $d\cA_\pm$
in \eqref{eq:Dintro} is nonzero at each puncture.
The theorem excludes the annulus ansatz with localized D7-branes
of Lozano, Macpherson and Montero \cite[\S3]{Lozano:AdS6Tduals} for all choices of its parameters.
On continuation around an orientifold puncture, the seven-brane transformation
of the two differentials is accompanied by an overall minus sign. With this
sign included in the global construction and the local asymptotics of
\cref{sec:O7}, \cref{thm:extended} again gives a disk, with at most one orientifold puncture.

The proof in the monodromy-free case compares two statements about the ratio
$\lambda=d\cA_+/d\cA_-$, which has unit modulus on $\partial\Sigma$. Let $N$ be the number of
zeros of $d\cA_+$ in the interior of $\Sigma$, counted with multiplicity, and let $2\pi d_i$ be
the increase of $\arg\lambda$ along the boundary component $\Gamma_i$, which carries $L_i$ of the
poles. Interior regularity excludes zeros of $d\cA_-$, so $N=\sum_i d_i$ by the argument
principle. Positivity and superharmonicity of a real function $\cG$ that enters the metric,
together with the pole asymptotics, bound the winding on each component. The count of zeros and
poles of $d\cA_+$ on the Schottky double fixes $N$ \cite[\S6.4, (6.23)]{DHoker:2017mds}. The
two results are
\begin{equation}
  d_i<L_i ,
  \qquad
  N=L+4g+2\nu-4 .
  \label{eq:localintro}
\end{equation}
Summing the first over the components gives $N<L$, and the second then gives \eqref{eq:main}.

Around a seven-brane puncture, monodromy can mix $d\cA_+$ and $d\cA_-$,
so their ratio $\lambda$ need not return to its original value.
For mutually local seven-branes, the common charge direction selects the
differential
\begin{equation}
  D\,dw=(\alpha\,\partial_w\cA_+-\bar\alpha\,\partial_w\cA_-)\,dw ,
  \label{eq:Dintro}
\end{equation}
where $\alpha$ is a constant of unit modulus fixed up to sign.
The global assumptions stated above make $D\,dw$ single-valued. Interior regularity
excludes zeros at regular points, and the assumption $D\ne0$ at the punctures
is the condition that each puncture carries a seven-brane source.
The zeros and poles of $D\,dw$ on the double can therefore be counted along the boundary,
where the phase rotation of $\lambda$ still obeys the bounds used in the
monodromy-free proof. Comparing this count with the bounds again forces $\Sigma$ to be a disk.
With orientifold punctures $D$ is defined only up to a sign, and the count is applied to its
square.

Sections~\ref{sec:setup}--\ref{sec:disk} give the regularity conditions,
boundary estimates and proof in the monodromy-free case.
\Cref{sec:punctures} treats seven-branes and orientifolds.
In \cref{sec:consequences} we also recover the charge ordering of Fluder and
Uhlemann \cite{Fluder:2020pym} from the boundary estimates.
We conclude with a discussion in \cref{sec:discussion}.
The counting steps of the three theorems and several of the local statements have been
checked in Lean~4; the scope is described in the AI statement at the end of the paper.

\section{The Type IIB \texorpdfstring{$\ads_6$}{AdS6} solutions}
\label{sec:setup}

We collect the properties of the Type IIB solutions used in the proof
\cite{DHoker:2016ujz,DHoker:2017mds,DHoker:2017zwj}.
The Riemann surface $\Sigma$ has genus $g$, $\nu$ boundary components and
a total of $L$ five-brane poles on its boundary. In the monodromy-free case,
$N$ denotes the number of zeros of $d\cA_+$ in the interior of $\Sigma$,
counted with multiplicity.

\subsection{Local solutions}
\label{sec:local}

The $SO(2,5)\times SO(3)$ isometry of \eqref{eq:ansatz} fixes the bosonic ansatz: the five-form
vanishes, the complex three-form is $F_{(3)}=d\cC\wedge\mathrm{vol}_{S^2}$, and $f_6$, $f_2$,
$\rho$, $\cC$ and the axion-dilaton $B$ are functions on $\Sigma$ alone. The BPS equations were solved completely in DGKU~I \cite{DHoker:2016ujz}: the general local solution is determined by two locally
holomorphic functions $\cA_\pm$ on $\Sigma$ through the two real combinations
\begin{equation}
  \kappa^2=|\partial_w\cA_-|^2-|\partial_w\cA_+|^2 ,
  \qquad
  \cG=|\cA_+|^2-|\cA_-|^2+\cB+\bar\cB ,
  \label{eq:kappaG}
\end{equation}
where the locally holomorphic $\cB$ is defined up to an additive constant by
\begin{equation}
  \partial_w\cB=\cA_+\partial_w\cA_--\cA_-\partial_w\cA_+ .
  \label{eq:Bdef}
\end{equation}
Under a holomorphic change of coordinate, $\kappa^2$ transforms so that
$\frac{i}{2}\kappa^2dw\wedge d\bar w$ is invariant; $\cG$ itself is a scalar. In
Einstein frame the metric factors are
\begin{equation}
  f_6^2=\frac{c_6^2\,\kappa^2(1+R)}{\rho^2(1-R)} ,
  \qquad
  f_2^2=\frac{c_6^2\,\kappa^2(1-R)}{9\rho^2(1+R)} ,
  \qquad
  \rho^2=\frac{(R+R^2)^{1/2}(\kappa^2)^{3/2}}{|\partial_w\cG|\,(1-R)^{3/2}} ,
  \label{eq:metricfactors}
\end{equation}
with $c_6$ a constant and $R$ determined by
\begin{equation}
  R+\frac1R=2+\frac{6\,\kappa^2\,\cG}{|\partial_w\cG|^2} ,
  \label{eq:Rdef}
\end{equation}
on the branch $0\le R<1$ that positivity of the metric factors selects
\cite[\S2.3]{DHoker:2017mds}. In the interior, $R=0$ occurs exactly at the critical
points of $\cG$, where the right-hand side of \eqref{eq:Rdef} diverges. The metric
factors remain finite and positive there. With $u:=(1-R)/(1+R)$ and
$c_6^2=1$, the normalization at a five-brane pole
\cite[(3.39)--(3.40)]{DHoker:2017mds}, they are
\begin{equation}
  f_6^2=\sqrt{6\cG}\,u^{-1/2} ,
  \qquad
  f_2^2=\tfrac19\sqrt{6\cG}\,u^{3/2} ,
  \qquad
  \rho^2=\frac{\kappa^2}{\sqrt{6\cG}}\,u^{-1/2} .
  \label{eq:factorsu}
\end{equation}
Two further properties will be used throughout: $\cG$ is superharmonic, and the duality
group acts linearly on the differentials. Differentiating
\eqref{eq:kappaG} and noting that $\cB+\bar\cB$ is harmonic gives
\begin{equation}
  \partial_w\partial_{\bar w}\cG=-\kappa^2 ,
  \label{eq:poisson}
\end{equation}
which is \cite[(2.14)]{DHoker:2017mds},
so $\cG$ is \emph{superharmonic}: its flat Laplacian
$\Delta=\partial_x^2+\partial_y^2=4\partial_w\partial_{\bar w}$ is negative wherever
$\kappa^2>0$, with $w=x+iy$. A non-constant superharmonic function cannot attain its
minimum in the interior. This is the minimum principle used below.

The classical Type IIB S-duality group acts on the differentials by $SU(1,1)$,
\begin{equation}
  \begin{pmatrix}\partial_w\cA_+\\ \partial_w\cA_-\end{pmatrix}
  \longmapsto
  M\begin{pmatrix}\partial_w\cA_+\\ \partial_w\cA_-\end{pmatrix} ,
  \qquad
  M=\begin{pmatrix}\mathsf{a}&\mathsf{b}\\ \bar{\mathsf{b}}&\bar{\mathsf{a}}\end{pmatrix} ,
  \quad |\mathsf{a}|^2-|\mathsf{b}|^2=1 ,
  \label{eq:su11}
\end{equation}
under which $\kappa^2$ and $\cG$ are invariant. A \emph{monodromy} is such an $M$ attached to a
closed circuit: continuing $(\partial_w\cA_+,\partial_w\cA_-)$ around a puncture brings it back
multiplied by $M$, and we write $M_j$ for the monodromy of the puncture $w_j$. The duality group is
also described as $SL(2,\mathbb R)$, acting on $\tau=\chi+ie^{-2\phi}$, with $\chi$ the axion and $\phi$ the dilaton, by
$\tau\mapsto(a\tau+b)/(c\tau+d)$ with real $a,b,c,d$ and $ad-bc=1$. The Cayley transform $B=(1+i\tau)/(1-i\tau)$ maps
the upper half-plane to the unit disk and turns this action into one of $SU(1,1)$ on $B$. In the
conventions of \cite[\S2.2, \S5.1]{DHoker:2016ujz}, the matrix $M$ in \eqref{eq:su11} is then
given by
\begin{equation*}
  \mathsf a=\tfrac12\big[(a+d)+i(b-c)\big] ,
  \qquad
  \mathsf b=\tfrac12\big[(a-d)-i(b+c)\big] .
\end{equation*}
Equivalently, the combinations $\partial_w\cA_++\partial_w\cA_-$ and
$i(\partial_w\cA_+-\partial_w\cA_-)$ form a doublet on which the real matrix with entries
$a,b,c,d$ acts.
We use $SU(1,1)$ for the holomorphic differentials and $SL(2,\mathbb R)$ for the brane
charges.

\subsection{Global regularity and five-brane charges}
\label{sec:global}

We require the metric factors to be real and positive in the interior of $\Sigma$,
so $\kappa^2>0$ and $\cG>0$ there. We consider surfaces whose completion is compact
after filling in the prescribed punctures and boundary poles. Such a surface must
have a boundary: otherwise, integrating the Poisson equation \eqref{eq:poisson}
and applying Stokes' theorem gives zero on one side and a strictly negative
integral on the other side. The same argument applies in the
presence of the prescribed seven-brane and orientifold punctures, since the
flux of $\cG$ through small circles around them vanishes.\footnote{%
At these punctures, $\cG$ is single-valued and bounded, and the integral of
$\kappa^2$ over a small disk is finite. To verify that the flux vanishes, let $\overline{\cG}(r)$ and
$\overline{\kappa^2}(r)$ denote angular averages near a puncture. Equation
\eqref{eq:poisson} gives $(r\overline{\cG}'(r))'=-4r\overline{\kappa^2}(r)$.
Integrating from $r$ to a fixed small radius $r_0$ gives
$r\overline{\cG}'(r)=r_0\overline{\cG}'(r_0)
+4\int_r^{r_0}s\overline{\kappa^2}(s)\,ds$.
Since the integral of $\kappa^2$ over the disk is finite, the right-hand side has a
finite limit $K$ as $r\to0$. If $K\ne0$, a further integration gives
$\overline{\cG}(r)=K\log r+o(|\log r|)$, contradicting boundedness of $\cG$
and hence of its angular average. Thus $K=0$, and the flux through the circle,
$\int_0^{2\pi}\partial_r\cG(r,\theta)\,r\,d\theta
=2\pi r\overline{\cG}'(r)$, tends to zero.
The local expressions are given in \cref{sec:sevenbranes,sec:O7}. At an orientifold,
integrating the $\mathcal O(r^{-1/2}\log r)$ differentials in \eqref{eq:O7local}
and \eqref{eq:Bdef} gives bounded $\cG$.}
We therefore denote the boundary components by $\Gamma_i$, $i=1,\ldots,\nu$,
with $\nu\ge1$, using the notation of DGU~II.

Along a regular boundary arc, the $S^2$ must shrink smoothly while the $\ads_6$
radius remains finite. Regularity therefore requires the holomorphic data to satisfy
\begin{equation}
  \kappa^2=\cG=0\ \text{on}\ \partial\Sigma .
  \label{eq:regularity}
\end{equation}
In a local coordinate $w=x+iy$ that places the boundary on the real axis and
$\Sigma$ in $y>0$, the differentials of the DGU~II class are taken to obey the
reflection condition
\begin{equation}
  \partial_w\cA_-=-\overline{\partial_w\cA_+}\qquad\text{on }\partial\Sigma .
  \label{eq:bdycond}
\end{equation}
These conditions apply on the open boundary arcs. Throughout,
physical regularity allows the prescribed localized brane singularities and
requires the supergravity fields to be regular away from the sources.

The reflection condition \eqref{eq:bdycond} ensures only that $\kappa^2=0$ and that
$\cG$ is constant on each regular boundary arc, as follows from the identities in
\cref{sec:identities}. Once these constants are required to agree across the five-brane poles, there is
one constant on each $\Gamma_i$, not yet set to zero. The common
additive constant in $\cG$ can be chosen to set its value on $\Gamma_1$ to zero.
The requirement $\cG=0$ in \eqref{eq:regularity} then imposes $\nu-1$ further
conditions, one for each remaining boundary component. As shown in
\cref{sec:identities}, $\cG$ also tends to zero at each five-brane pole from
the interior. At the prescribed interior punctures $\cG$ is bounded and
extends across them as a superharmonic function. The minimum principle and
\eqref{eq:poisson} then give $\cG>0$ in the regular interior, so $\cG$ attains
its global minimum on every boundary component, as used in \cref{lem:rotation}.

Away from the prescribed interior punctures, regularity excludes poles of
$\partial_w\cA_\pm$ in the interior. At the candidate interior pole analyzed in
DGU~II, $\cA_-$ diverges logarithmically while $\cA_+$ remains finite, so
$\cG\sim-|\cA_-|^2\to-\infty$, in conflict with $\cG>0$
\cite[App.~A, (A.8)]{DHoker:2017mds}. The same conclusion holds at any interior pole.
Since $\kappa^2>0$, the ratio $\lambda=\partial_w\cA_+/\partial_w\cA_-$ is holomorphic
with $|\lambda|<1$ in a punctured neighborhood, so it extends to the pole, where
$|\lambda|<1$ by the maximum principle. Thus $\partial_w\cA_-$ is singular there and
$\partial_w\cA_+=\lambda\,\partial_w\cA_-$. To leading order $\cA_+\simeq\lambda\,\cA_-$
with $\lambda$ evaluated at the pole, while $\cB$ is subleading, so
$\cG\simeq-(1-|\lambda|^2)|\cA_-|^2\to-\infty$. The five-brane poles therefore
lie on $\partial\Sigma$. We denote their total number by $L$, their positions by
$p_\ell$, which we call the five-brane points, and their residues by $Z^\ell_\pm$, with $\ell=1,\ldots,L$; $L_i$ counts the poles on $\Gamma_i$,
so $L=\sum_i L_i$, as illustrated in \cref{fig:setup}(a). The reflection condition
\eqref{eq:bdycond} implies
\begin{equation}
  Z_-^\ell=-\overline{Z_+^\ell} ,
  \label{eq:residues}
\end{equation}
regardless of the topology or the presence of punctures. Although derived in a real
boundary coordinate, this relation holds in any holomorphic coordinate, since
$Z^\ell_\pm$ are the coordinate-invariant residues of the one-forms $d\cA_\pm$.
We abbreviate
$Z_\ell=Z^\ell_+$.
At each boundary pole the Type IIB supergravity fields have the near-horizon
asymptotics of a $(p,q)$ five-brane source. The charge of the corresponding
external five-brane stack is determined by the residue $Z_\ell$
\cite[\S3.9--3.10]{DHoker:2017mds}. A regular solution has at least one
pole on each boundary component. In the monodromy-free case this follows
from the boundary winding argument in \cref{sec:disk}, and with
interior punctures from \cref{lem:polefree} under the assumptions stated there.

In the monodromy-free class, the charges also satisfy
\begin{equation}
  \sum_{\ell=1}^{L}Z^\ell_\pm=0
  \qquad\text{(single-valued differentials, no punctures)} ,
  \label{eq:chargesum}
\end{equation}
by the residue theorem on the double. The Schottky double $\hat\Sigma$ is formed by gluing $\Sigma$ to
its mirror, equipped with the conjugate complex structure, along their common boundary.
Its genus is $\hat g=2g+\nu-1$, where $g$ is the genus of $\Sigma$.
Schwarz reflection using \eqref{eq:bdycond} then extends the single-valued differentials
meromorphically across this boundary, and their total residues on the resulting closed
surface must vanish. Each boundary pole is a single point on the double and retains its residue
$Z^\ell_\pm$, so it is counted only once. The sum includes poles on all boundary components and does not impose charge
conservation separately on each one.\footnote{For $\hat g>0$, the differentials may also
contain globally holomorphic one-forms on the double, which change neither their
poles nor their residues. On the annulus these include the constant one-forms of
\cite[\S5]{DHoker:2017mds}. The functions $\cA_\pm$ may have additive periods, which are
allowed in the argument; see \cref{lem:polefree-bound}.}
With seven-brane punctures, the external five-brane charges need not sum to zero
by themselves; charge conservation also involves the seven-brane sector, with
charges compared in a common duality frame \cite{DHoker:2017zwj}.

For an unpunctured disk the solution is explicit.
In the upper half-plane representation, the holomorphic functions are exactly
$\cA_\pm(w)=\cA^0_\pm+\sum_\ell Z^\ell_\pm\ln(w-p_\ell)$,
where $p_\ell\in\mathbb R$ and $\cA^0_\pm$ are complex integration constants.
The double is a sphere, so there are no additional globally holomorphic one-forms.
The boundary values of $\cG$ must match across the five-brane poles
\cite[\S3.6]{DHoker:2017mds}. Writing $\cA^0:=\tfrac12(\cA^0_+-\overline{\cA^0_-})$, this
gives the conditions
\begin{equation}
    \cA^0Z^\ell_-+\overline{\cA^0}\,Z^\ell_++\sum_{m\ne \ell}Z^{[m,\ell]}\ln|p_m-p_\ell|=0 ,
  \qquad
  Z^{[m,\ell]}:=Z^m_+Z^\ell_--Z^\ell_+Z^m_- ,
  \label{eq:regcond}
\end{equation}
for $\ell=1,\dots,L$ \cite[(3.25)]{DHoker:2017mds}.
DGU~II uses the gauge $\cA^0_-=-\overline{\cA^0_+}$, in which
$\cA^0=\cA^0_+$; the other combination of integration constants changes only
the gauge of the two-form potential.\footnote{For a complex constant $c$, the shifts
$\cA_+\mapsto\cA_++c$, $\cA_-\mapsto\cA_-+\bar c$ leave $\cA^0$ unchanged.
Together with $\cB\mapsto\cB+c\cA_--\bar c\cA_+$, they preserve $\kappa^2$ and
$\cG$, hence the metric. The axion--dilaton is also unchanged, while $\cC$ shifts
by a constant, leaving $F_{(3)}=d\cC\wedge\mathrm{vol}_{S^2}$ invariant.
This is the two-form gauge transformation in DGKU~I \cite[(5.18)]{DHoker:2016ujz}.}

The expression above is written in terms of the poles and residues, and does not display
the zeros of the differentials. The one-form $d\cA_+$ has $L$ simple poles on the sphere and is
regular at infinity by \eqref{eq:chargesum}, so it has $L-2$ zeros $s_n$. DGU~II take the poles and these zeros as the
parameters \cite[\S3.1]{DHoker:2017mds},
\begin{equation}
  \partial_w\cA_+=\omega_0\lambda_0\,\frac{\prod_{n=1}^{L-2}(w-s_n)}{\prod_{\ell=1}^{L}(w-p_\ell)} ,
  \qquad
  \partial_w\cA_-=\omega_0\bar\lambda_0\,\frac{\prod_{n=1}^{L-2}(w-\bar s_n)}{\prod_{\ell=1}^{L}(w-p_\ell)} ,
  \label{eq:zeroparam}
\end{equation}
with $|\lambda_0|=1$ and $\bar\omega_0=-\omega_0$, which solves \eqref{eq:bdycond}. The residues
$Z^\ell_\pm$ are then determined by the poles and zeros. Since $|w-\bar s_n|>|w-s_n|$ when $w$
and $s_n$ both lie in the upper half-plane, placing all the zeros $s_n$ in the interior gives
$\kappa^2>0$ there \cite[\S3.2]{DHoker:2017mds}. At least one zero is needed: for $L=2$ the
differentials \eqref{eq:zeroparam} have equal modulus and $\kappa^2$ vanishes identically, so a
solution without punctures has $L\ge3$ \cite[\S3.3]{DHoker:2017mds}. Since $\cG$ vanishes on the boundary arcs and
tends to zero at the poles, $\cG>0$ then follows from the minimum principle. On the disk the conditions \eqref{eq:regcond}, imposed on the parameters
of \eqref{eq:zeroparam}, are therefore sufficient as well as necessary for a regular solution.

This freedom is special to the disk. On a surface of any other topology the double has
positive genus, and the number of zeros of $d\cA_+$ on the double exceeds $L-2$ by twice
that genus. Positivity of $\kappa^2$ still requires all of them to lie in $\Sigma$ and none in
its mirror. The theorems below show that this is incompatible with the boundary behavior
of $\lambda$.

Charge conservation and the antisymmetry of $Z^{[m,\ell]}$ imply that the sum of the
$L$ equations \eqref{eq:regcond} vanishes identically, so at most $L-1$ of them are independent.
The parametrization \eqref{eq:zeroparam} has $3L-2$ real parameters, equivalent to
the $L$ pole positions and the $L$ residues $Z_\ell$, which obey \eqref{eq:chargesum}. The group
$PSL(2,\mathbb R)$ moves the poles and leaves the residues unchanged, so $L-3$ cross-ratios of
the poles remain. Together with the complex constant $\cA^0$ these are $L-1$ real unknowns for
the $L-1$ equations. If the equations are independent, the solutions for given residues are
isolated, and the $2L-2$ real parameters that remain are the independent charges of the conformal
five-brane web \cite[(3.28), \S3.8]{DHoker:2017mds}. Existence and uniqueness are discussed in
\cref{sec:legs}.

The topology argument uses the local boundary
properties established in \cref{sec:boundary}; its extension to punctured surfaces
requires the additional global assumptions introduced next.

\subsection{Seven-branes and orientifolds}
\label{sec:sevenbranes}

Following DGU~III \cite{DHoker:2017zwj}, we restrict the seven-brane punctures at
$w=w_j$ to be mutually local: their charges share a common $[p,q]$ direction,
and their monodromies commute. For coprime integers $p,q$, a single
seven-brane has monodromy
$M_{[p,q]}=\left(\begin{smallmatrix}1-pq&p^2\\-q^2&1+pq\end{smallmatrix}\right)
\in SL(2,\mathbb Z)$.
A puncture $w_j$ contains $m_j$ branes of this same charge, where $m_j$ is a
positive integer, and has monodromy $M_j=M_{[p,q]}^{m_j}$. Since
$(M_{[p,q]}-\mathbf1)^2=0$, this gives
$M_j=\mathbf1+m_j(M_{[p,q]}-\mathbf1)$.

We use the notation $m_j=n_j^2$ of DGU~III, with $n_j$ real, and call $n_j^2$ the strength
of the puncture. In the matrix
parametrization of DGU~III, $M_{[p,q]}^{m_j}=M_{[n_jp,n_jq]}$
\cite[(3.1),(3.15)]{DHoker:2017zwj}. The pair $[n_jp,n_jq]$ parametrizes the
monodromy matrix of the stack, not the charge of an individual brane.
Each constituent has charge $[p,q]$, and $n_j^2$ can be any positive integer.\footnote{At the classical
supergravity level, the construction also allows continuous non-negative
strengths $n_j^2$ and $SL(2,\mathbb R)$ monodromies. At zero strength
the seven-brane source is absent.}
A single duality transformation sends the common charge direction to $[1,0]$.
In this D7-brane frame, all punctures have monodromy $M_j=T^{n_j^2}$, where
$T=\left(\begin{smallmatrix}1&1\\0&1\end{smallmatrix}\right)$.
The transformation leaves $\kappa^2$, $\cG$ and the boundary regularity conditions unchanged.
Each puncture thus carries a \emph{parabolic} monodromy, $|\mathrm{tr}\,M_j|=2$,
which preserves a charge direction. Composite seven-branes with \emph{elliptic} monodromy,
$|\mathrm{tr}\,M_j|<2$, such as those of type $E_{6,7,8}$ and $H_{0,1,2}$
\cite{Apruzzi:Sfolds}, preserve none and are not included, even when only one puncture is
present.

DGU~III constructs disk solutions in the D7-brane frame by adding monodromy
to a pair of seed functions. The probe limit,
brane-web interpretation and sphere partition functions of these solutions were studied in
\cite{Gutperle:2018vdd}, and further solutions were obtained in
\cite{Chaney:2018gjc}. In the $SU(1,1)$ basis of \eqref{eq:su11}, the D7-brane
monodromy $T^{n_j^2}$ acts on the pair $(d\cA_+,d\cA_-)$ by the matrix
\[
  \mathbf 1+\frac{i n_j^2}{2}\begin{pmatrix}1&-1\\1&-1\end{pmatrix}.
\]
Thus the monodromy shifts both differentials by
$\tfrac{i n_j^2}{2}(d\cA_+-d\cA_-)$, leaving their difference invariant.
A function $f$ that shifts by $i n_j^2/2$ on a counterclockwise
circuit of $w_j$ implements this monodromy. We write the disk construction
of DGU~III \cite[(3.8), (3.11)]{DHoker:2017zwj} in terms of the Dirichlet
Green's function as
\begin{equation}
  \partial_w\cA_\pm=\partial_w\cA^{\mathrm{seed}}_\pm
  +f\,\big(\partial_w\cA^{\mathrm{seed}}_+-\partial_w\cA^{\mathrm{seed}}_-\big) ,
  \qquad
  \mathrm{Re}\,f=-\frac1{4\pi}\sum_j n_j^2\,\mathsf{g}(w,w_j) .
  \label{eq:sevenbrane}
\end{equation}
Here $\mathsf g$ is the Dirichlet Green's function on $\Sigma$, normalized
by $\Delta_w\mathsf g(w,w_j)=-2\pi\delta^{(2)}(w-w_j)$. It vanishes on
$\partial\Sigma$ and behaves as $\ln(1/|w-w_j|)$ near $w_j$.
The functions $\cA^{\mathrm{seed}}_\pm(w)$ are the monodromy-free seed functions of DGU~III,
written there as $\cA^{(0)}_\pm(w)$. On the disk they have the same
constant-plus-logarithm structure as in \cref{sec:global}. The term containing $f$
introduces the seven-brane monodromies and also changes the boundary residues.
Writing $Z_{\pm,\mathrm{seed}}^\ell$ for the seed residues, the full residues
$Z_\pm^\ell$ obey, in a branch where $f$ is regular near $p_\ell$,
\[
  Z_\pm^\ell=Z_{\pm,\mathrm{seed}}^\ell
  +f(p_\ell)\bigl(Z_{+,\mathrm{seed}}^\ell-Z_{-,\mathrm{seed}}^\ell\bigr).
\]
Thus even on the disk, prescribed external five-brane charges constrain this
combination of seed and puncture data, rather than the seed residues alone.

Lozano, Macpherson and Montero adapted the disk construction of
DGU~III to the annulus. For a given annulus modulus and puncture data, they
expressed the monodromy function $f$ in terms of theta functions
\cite[(3.1)]{Lozano:AdS6Tduals}. They left open whether the parameters of
this ansatz could be chosen to satisfy the regularity conditions on both
boundary components simultaneously \cite[\S\S3, 9]{Lozano:AdS6Tduals}.
In their ansatz the seed differentials are single-valued, and the shift
$f(w+1)-f(w)$ accounts for the full monodromy of $d\cA_\pm$ around the annulus
\cite[(A.10), (A.13)--(A.14)]{Lozano:AdS6Tduals}.

On a surface of arbitrary topology we assume the same global construction
\eqref{eq:sevenbrane}, with one pair of single-valued meromorphic seed
differentials satisfying \eqref{eq:bdycond} and having poles only at the
boundary five-brane points. The Dirichlet Green's function fixes
$\mathrm{Re}\,f$, so the Cauchy--Riemann equations determine the change in
$\mathrm{Im}\,f$ along any closed curve in the punctured interior. We assume
that the resulting period accounts for the full duality transformation of
$d\cA_\pm$ along the curve. This is the assumption of no independent
duality holonomy used in \cref{lem:shift}.

In the D7-brane frame, the seven-brane monodromies leave $D\,dw=d\cA_+-d\cA_-$ invariant,
corresponding to $\alpha=1$ in \eqref{eq:Dintro}.
By \eqref{eq:sevenbrane} the terms containing $f$ cancel in this difference, so
$D=\partial_w\cA_+^{\mathrm{seed}}-\partial_w\cA_-^{\mathrm{seed}}$ is fixed by the seed alone
and is finite at the punctures $w_j$.
At a five-brane pole $p_\ell$ its residue is
$Z_+^\ell-Z_-^\ell=2\operatorname{Re}Z_+^\ell$ by \eqref{eq:residues},
and is proportional to the NS5 charge \cite[(5.1)]{DHoker:2017zwj}.
With $\xi=w-w_j$,
the behavior of $\kappa^2$ near a seven-brane puncture is
\begin{equation}
  \kappa^2=-\frac{n_j^2}{4\pi}\,|D(w_j)|^2\,\ln|\xi|^2+\mathcal{O}(1) ,
  \label{eq:sevenkappa}
\end{equation}
The coefficient $c$ of DGU~III \cite[(3.64)]{DHoker:2017zwj} equals
$-D(w_j)$ in our convention. Define the seed quantity
$\kappa^2_{(0)}:=|\partial_w\cA_-^{\mathrm{seed}}|^2-|\partial_w\cA_+^{\mathrm{seed}}|^2$.
If $\kappa^2_{(0)}>0$ at $w_j$, as in the disk construction of
DGU~III \cite[(3.6)]{DHoker:2017zwj}, the seed differentials have unequal
magnitudes there, so their difference $D(w_j)$ is nonzero. On a general surface
we impose $D(w_j)\ne0$ as a local source condition. This excludes little.
By \eqref{eq:sevenbrane},
\begin{equation}
  \kappa^2=\kappa^2_{(0)}-2\,\mathrm{Re}f\,|D|^2 ,
  \qquad
  \kappa^2_{(0)}=-2\,\mathrm{Re}\big(\overline{\partial_w\cA_+^{\mathrm{seed}}}\,D\big)+|D|^2 .
  \label{eq:kappaseed}
\end{equation}
If $D$ vanishes at $w_j$ while $\partial_w\cA_+^{\mathrm{seed}}$ does not, the first term of
$\kappa^2_{(0)}$ dominates near $w_j$ and is negative in some directions, so regularity
already fails. The condition therefore excludes a possibly regular solution only if both
seed differentials vanish at the puncture. In that case $\kappa^2$ has no logarithmic
divergence, the expansion of $R$ in \cite[(3.68)]{DHoker:2017zwj} does not apply, and the
local geometry is not that of the seven-brane of DGU~III. We do not consider this case.
For $D(w_j)\ne0$ the local analysis gives $\cG=G_j+\mathcal{O}(|\xi|)$ with a finite constant $G_j>0$.

The identity \eqref{eq:kappaseed} also shows that the bound $L\ge3$ of \cref{sec:global} is
special to solutions without punctures. For a seed with $L=2$, $\kappa^2_{(0)}$ vanishes
identically and $D$ has no zeros unless it vanishes identically. Then
$\kappa^2=-2\,\mathrm{Re}f\,|D|^2$ is positive in the interior, because $\mathrm{Re}f<0$ there.
These are the two-pole solutions of Chaney and Uhlemann \cite[\S3.1]{Chaney:2018gjc}.

An orientifold puncture carries monodromy $-T^k$, with $k=n_j^2$
\cite{Uhlemann:2019lge}. The extra minus sign reverses both seed differentials
on continuation around the puncture, so $D\,dw$ also changes sign. This holds
even for $k=0$, when the monodromy is the central element $-\mathbf{1}$.
We give the local expression in \cref{sec:O7}.

\section{Boundary regularity and phase rotation}
\label{sec:boundary}

In this section we bound the rotation of the phase of the ratio
$\lambda=\partial_w\cA_+/\partial_w\cA_-$ of DGU~II \cite[(2.17)]{DHoker:2017mds} along a
boundary component $\Gamma_i$. Positivity of $\kappa^2$ fixes the direction in which this
phase rotates, while positivity and superharmonicity of $\cG$ fix the sign of its inward
normal derivative. Combining these properties with the pole asymptotics gives
the bound $d_i<L_i$ of \eqref{eq:localintro}.

\subsection{Boundary phase and pole asymptotics}
\label{sec:identities}

We first establish monotonicity of the boundary phase, continuity of $\cG$ at
the five-brane poles, and an identity for its inward normal derivative.

On every component $\Gamma_i$ of $\partial\Sigma$, the reflection condition
\eqref{eq:bdycond} gives
\begin{equation}
  |\lambda|=1 ,
  \qquad
  \arg\lambda=\pi+2\arg\partial_w\cA_+ ,
  \label{eq:lambdabdy}
\end{equation}
where phases are chosen continuously along each arc between consecutive poles.
We show below that $\partial_w\cA_+$ does not vanish on these arcs.
We traverse the boundary \emph{positively}, with $\Sigma$ on the left; in a local
upper-half-plane coordinate this means increasing $x$, with $\partial_y$
pointing inward. Figure~\ref{fig:setup} fixes these conventions.

\begin{figure}[t]
\centering
\begin{tikzpicture}[scale=1.05,>=latex]
\begin{scope}
  \fill[gray!14] (0,0) circle (1.75);
  \fill[white]   (0.45,-0.25) circle (0.70);
  \draw[thick]   (0,0) circle (1.75);
  \draw[thick]   (0.45,-0.25) circle (0.70);
  \draw[->,very thick] (10:2.02) arc (10:48:2.02);
  \begin{scope}[shift={(0.45,-0.25)}]
    \draw[->,very thick] (-15:0.88) arc (-15:-75:0.88);
  \end{scope}
  \foreach \a in {105,150,205,330} \fill (\a:1.75) circle (2.4pt);
  \begin{scope}[shift={(0.45,-0.25)}]
    \foreach \a in {75,195} \fill (\a:0.70) circle (2.4pt);
  \end{scope}
  \draw[dashed] (-0.72,0.34) -- (177:1.75);
  \node at (-0.72,0.34) {$\times$};
  \node at (-0.72,0.66) {\small $w_j$};
  \node at (-1.20,-0.02) {\small cut};
  \node at (0.62,1.08) {$\Sigma$};
  \node at (128:2.06) {\small $\Gamma_1$};
  \begin{scope}[shift={(0.45,-0.25)}]
    \node at (147:0.40) {\small $\Gamma_2$};
  \end{scope}
  \node at (330:2.06) {\small $p_\ell$};
  \node at (0,-2.45) {(a)};
\end{scope}
\begin{scope}[xshift=6.5cm]
  \fill[gray!14] (-2.0,0) rectangle (2.0,1.40);
  \draw[thick] (-2.32,0) -- (2.32,0);
  \fill (-1.10,0) circle (2.4pt);
  \fill ( 1.00,0) circle (2.4pt);
  \node at (-1.10,-0.32) {\small $p_\ell$};
  \node at ( 1.00,-0.32) {\small $p_{\ell+1}$};
  \draw[very thick] (-1.00,0) -- (0.90,0);
  \node at (-0.55,0.24) {\small open arc};
  \draw[->,thick] (0.55,0.12) -- (0.55,0.88);
  \node[anchor=west] at (0.62,0.60) {\small $\partial_n=\partial_y$};
  \draw[->,very thick] (-0.45,-0.74) -- (0.45,-0.74);
  \node at (0.86,-0.74) {\small $+x$};
  \node at (-1.45,1.02) {$\Sigma$};
  \node at (0,-2.45) {(b)};
\end{scope}
\end{tikzpicture}
\caption{(a) A surface with two boundary components, five-brane poles (dots), and a
seven-brane puncture ($\times$). The dashed line is a branch cut. The holomorphic data jump
across it, while $\kappa^2$ and $\cG$ remain invariant. The cut direction and its position
relative to the poles enter the construction \cite[\S3.6, \S4]{DHoker:2017zwj}; cuts are
chosen to meet neither poles nor one another \cite[\S3.3]{DHoker:2017zwj}. Positive boundary
orientation keeps $\Sigma$ on the left. (b) Local coordinates near a boundary arc, with
$\Sigma$ in the upper half-plane and the inward normal pointing along $+y$. For disk topology,
the upper half-plane also provides a global conformal representation.}

\label{fig:setup}
\end{figure}

Since $\kappa^2>0$ in the interior, $\partial_w\cA_-$ has no zeros there and
each local branch of $\lambda$ is holomorphic with $|\lambda|<1$.
If either $\partial_w\cA_+$ or $\partial_w\cA_-$ vanished at a point on an open
boundary arc, \eqref{eq:bdycond} would give $\partial_w\cA_+=\partial_w\cA_-=0$
there. Differentiating \eqref{eq:kappaG} would then give $\partial_w\cG=0$
at that point. This contradicts Hopf's boundary point lemma
\cite{Hopf:1952,GilbargTrudinger:1983} applied to the positive function
$\cG$, which vanishes on the arc and has strictly positive inward normal derivative.
Thus $\lambda$ is well defined and has unit modulus on every regular boundary
arc, as in DGU~II \cite[\S2.4.1]{DHoker:2017mds}. It is nonzero in a
sufficiently small neighborhood of each regular boundary point, so
$\log|\lambda|$ is harmonic there,
negative inside and zero on the boundary. Hopf's lemma gives a strictly negative inward normal derivative.
In the local upper-half-plane coordinate, a small disk tangent to the real axis
from above at the boundary point provides the interior ball required by Hopf's lemma.

Writing $\log\lambda=\log|\lambda|+i\psi$, the Cauchy--Riemann equations give
$\partial_x\psi=-\partial_y\log|\lambda|$. Take $s$ to be positively
oriented boundary arclength in any smooth conformal metric on $\Sigma$,
and $\partial_n$ the inward unit normal in the same metric. The two
coordinate derivatives acquire the same positive factor, giving
\begin{equation}
  \partial_s\psi=-\partial_n\log|\lambda|>0 .
  \label{eq:psimono}
\end{equation}
Since $\log|\lambda|$ is constant along the boundary,
$\partial_s\log\lambda=i\,\partial_s\psi$ there, that is
\begin{equation}
  |\partial_s\lambda|=\partial_s\psi>0 ,
  \label{eq:lambdaprime}
\end{equation}
so $\lambda':=\partial_w\lambda$ has no zero on a regular boundary arc.\footnote{On the
unpunctured disk this can be checked directly. By \eqref{eq:zeroparam},
$\lambda=\lambda_0^2\prod_{n=1}^{L-2}(w-s_n)/(w-\bar s_n)$, with $|\lambda_0|=1$ and
$\mathrm{Im}\,s_n>0$. On the boundary, the real axis, the $n$-th
factor has phase $2\arg(w-s_n)$, which increases monotonically. The argument in the text needs
no explicit expression for $\lambda$ and applies to any topology.}
In \cref{sec:count} we use $\lambda'\ne0$ to show that the boundary zeros of
the monodromy-invariant one-form $D\,dw$ in \eqref{eq:Dintro} are simple.

We record the increase of $\arg\lambda$ along $\Gamma_i$, traversed positively and measured
in units of $2\pi$, as
\begin{equation}
  d_i:=\frac1{2\pi}\,\Delta_{\Gamma_i}\arg\lambda\;>\;0 .
  \label{eq:ddef}
\end{equation}
In the monodromy-free case, $\lambda$ is single-valued. Since $\kappa^2>0$,
its denominator $\partial_w\cA_-$ never vanishes in the interior, so $\lambda$
has no interior poles and its zeros are precisely those of $\partial_w\cA_+$,
with the same multiplicities. At each boundary pole the simple poles in the
ratio cancel, leaving a finite nonzero value. The argument principle applies
on $\Sigma$ of any genus. Indeed, $d\lambda/\lambda$ is a globally defined
one-form, closed away from the zeros of $\lambda$. Removing small disks
around these zeros and applying Stokes' theorem gives
\[
  \sum_i d_i=\frac{1}{2\pi i}\int_{\partial\Sigma}\frac{d\lambda}{\lambda}
  =\sum_{\lambda(\zeta)=0}\operatorname{ord}_\zeta(\lambda),
\]
since the integral around each small positively oriented circle is $2\pi i$
times the order of the zero.
Each $d_i$ is the winding of $\lambda$ along $\Gamma_i$ with its induced
boundary orientation, and the sum over all components gives the total number
$N$ of interior zeros of $d\cA_+$, counted with multiplicity. We summarize these statements.

\begin{lemma}[Boundary winding]
\label{lem:degrees}
$d_i>0$ on every positively traversed boundary component. If $\lambda$ returns
to itself along $\Gamma_i$, then $d_i$ is a positive integer. In the
monodromy-free class, $\sum_i d_i$ equals the number of interior zeros of
$\partial_w\cA_+$, counted with multiplicity.
\end{lemma}

With punctures, continuation around $\Gamma_i$ multiplies the differentials by a monodromy
matrix $M_{\Gamma_i}$. The continuation takes the initial value $\lambda_{\rm in}$ to
$(\mathsf a\lambda_{\rm in}+\mathsf b)/(\bar{\mathsf b}\lambda_{\rm in}+\bar{\mathsf a})$, with
$\mathsf a$, $\mathsf b$ the entries of $M_{\Gamma_i}$ as in \eqref{eq:su11}. In a fixed duality frame the same transformation
acts at each starting point, but its angular displacement on the unit circle
can vary with $\lambda_{\rm in}$. Thus $d_i$ can depend on the starting point and
need not be an integer.
On a component containing poles we start at any five-brane pole, so that the
circuit consists of complete arcs from one pole to the next. The global count of
\cref{sec:count} is independent of this choice. We give the dependence of $d_i$
on a general base point in \cref{lem:lift}.

To determine the limit of $\cG$ at a five-brane pole $p_\ell$, we use the
local expansion of DGU~II \cite[\S3.6]{DHoker:2017mds}. With the residue
relation \eqref{eq:residues}, the holomorphic data are
\begin{equation}
  \cA_+=Z_\ell\log(w-p_\ell)+a_++\mathcal{O}(w-p_\ell) ,
  \qquad
  \cA_-=-\overline{Z_\ell}\log(w-p_\ell)+a_-+\mathcal{O}(w-p_\ell) .
  \label{eq:poleexp}
\end{equation}
Here $a_\pm$ are the finite parts at the chosen pole.
The logarithm-squared terms cancel in $|\cA_+|^2-|\cA_-|^2$. In
$\partial_w\cB$ of \eqref{eq:Bdef}, the terms proportional to $\log(w-p_\ell)/(w-p_\ell)$ also
cancel, leaving a simple pole of residue $-(a_+\overline{Z_\ell}+a_-Z_\ell)$. Let $G_\ell^{\rm R}$ denote
the limit of $\cG$ along the boundary arc to the right of $p_\ell$.
Substitution in \eqref{eq:kappaG} gives
\begin{equation}
\begin{aligned}
  \cG(w)&=G_\ell^{\rm R}-4\,C_\ell\,\arg(w-p_\ell)+\mathcal{O}\big(|w-p_\ell|\log|w-p_\ell|\big) ,\\
  C_\ell&=\mathrm{Im}\big(\bar a_+Z_\ell\big)+\mathrm{Im}\big(\bar a_-\overline{Z_\ell}\big) .
\end{aligned}
  \label{eq:contpole}
\end{equation}
On the boundary arcs to the right and left of $p_\ell$, the argument is $0$
and $\pi$, respectively. Since $\cG=0$ on both arcs, the right limit gives
$G_\ell^{\rm R}=0$ and the left limit then gives $C_\ell=0$. Thus only the
$\mathcal{O}(r\log r)$ term remains, and $\cG\to0$ also as $w$ approaches
$p_\ell$ from the interior, with $r=|w-p_\ell|\to0$.

To express the inward normal derivative of $\cG$ in terms of the holomorphic
data, define on each boundary component
\begin{equation}
  \hat\cA:=\tfrac12\big(\cA_+-\overline{\cA_-}\big) .
  \label{eq:Ahat}
\end{equation}
It is finite on each open arc between successive poles and diverges logarithmically at the
poles. Differentiating \eqref{eq:kappaG} and using \eqref{eq:bdycond} to eliminate $\partial_w\cA_-$
gives, on a regular boundary arc,
\begin{equation}
  \partial_w\cG
  =\big(\overline{\cA_+}-\cA_-\big)\partial_w\cA_+-\big(\cA_+-\overline{\cA_-}\big)
    \overline{\partial_w\cA_+}
  =2\,\overline{\hat\cA}\,\partial_w\cA_+-2\,\hat\cA\,\overline{\partial_w\cA_+} ,
\end{equation}
that is
\begin{equation}
  \partial_y\cG=-8\,\mathrm{Im}\big(\overline{\hat\cA}\,\partial_w\cA_+\big) ,
  \label{eq:dyg}
\end{equation}
with $\partial_y$ along the inward normal. On each regular boundary arc,
\eqref{eq:bdycond} also gives $\partial_x\hat\cA=\partial_w\cA_+$.

\subsection{The rotation bound}
\label{sec:rotation}

As above, let $s$ denote positively oriented boundary arclength.
Write $\dot{\hat\cA}=d\hat\cA/ds$ and set $P(s)=\overline{\hat\cA}\,\dot{\hat\cA}$
at regular boundary points. In a local upper-half-plane coordinate $w=x+iy$,
$\dot{\hat\cA}=(dx/ds)\,\partial_w\cA_+$ on the boundary. Since $dx/ds>0$,
these two quantities have the same phase. On each open arc $I$ between
successive poles, let $\delta_I=\Delta_I\arg\dot{\hat\cA}$.
By \eqref{eq:lambdabdy}, the phase of $\lambda$ increases by $2\delta_I$
on $I$. At a pole the common simple poles of $\partial_w\cA_\pm$ cancel
in their ratio, so $\lambda$ extends through it with nonzero value.
Thus $\sum_I\delta_I=\pi d_i$ for a circuit starting at a pole.

\begin{lemma}[Rotation bound]
\label{lem:rotation}
Suppose the value of $\cG$ along $\Gamma_i$ is its global minimum on
$\overline{\Sigma}$, and let $L_i\ge1$. In the absence of monodromy
the base point for $d_i$ is arbitrary.
With monodromy, start at any pole so that the circuit consists of complete
arcs between successive poles. Then $\delta_I<\pi$ on each arc $I$, and
\begin{equation}
  \sum_{I\subset\Gamma_i}\Delta_I\arg\hat\cA=\pi\,(d_i-L_i)<0 ,
  \qquad\text{so}\qquad d_i<L_i .
  \label{eq:rotation}
\end{equation}
\end{lemma}

\begin{proof}
By \eqref{eq:poisson}, $\cG$ is strictly superharmonic in the regular
interior. Since $\cG$ takes its minimum value along $\Gamma_i$,
at each regular point of this boundary component
Hopf's lemma gives $\partial_y\cG>0$ in the local upper-half-plane
coordinate. By \eqref{eq:dyg},
\begin{equation}
  \mathrm{Im}\,P=-\tfrac18(dx/ds)\,\partial_y\cG<0 .
  \label{eq:Qneg}
\end{equation}
In particular $\hat\cA$ is nonzero on each open arc, and
$d(\arg\hat\cA)/ds=\mathrm{Im}\,P/|\hat\cA|^2<0$ there.

Consider one arc $I$ traversed from pole to pole.
Near either endpoint $p$ of residue $Z$, write $x_p$ for its position in
the local boundary coordinate $x$. By \eqref{eq:poleexp} and
\eqref{eq:Ahat}, the two logarithms in $\hat\cA$ combine to give
$\hat\cA=Z\log|x-x_p|+\mathcal O(1)$, while
$\partial_x\hat\cA=Z/(x-x_p)+\mathcal O(1)$. Hence
\[
  P\sim \frac{dx}{ds}\,|Z|^2\,\frac{\log|x-x_p|}{x-x_p}.
\]
The logarithm is negative near either pole. The denominator is positive
at the start of $I$ and negative at the end. Since $P$ stays in the open
lower half-plane, its argument in $(-\pi,0)$ therefore tends to $-\pi$
at the start and to $0$ at the end. Thus $\Delta_I\arg P=\pi$.
Since conjugation reverses the phase, $P=\overline{\hat\cA}\,\dot{\hat\cA}$
gives, for phases followed continuously along $I$,
\[
  \pi=\Delta_I\arg P
     =\Delta_I\arg\dot{\hat\cA}-\Delta_I\arg\hat\cA
     =\delta_I-\Delta_I\arg\hat\cA .
\]
Thus $\Delta_I\arg\hat\cA=\delta_I-\pi$, which is negative because
$d(\arg\hat\cA)/ds<0$ on $I$. Summing over the $L_i$ arcs gives \eqref{eq:rotation}.
\end{proof}

For a physically regular solution, every boundary component attains the minimum
of $\cG$, so the rotation bound applies to every component containing poles.
The following estimate covers components without poles. It does not require
$\hat\cA$ to be single-valued along the component.

\begin{lemma}[Pole-free boundary bound]
\label{lem:polefree-bound}
Suppose the value of $\cG$ along $\Gamma_i$ is its global minimum on $\overline{\Sigma}$.
If $L_i=0$, then $d_i<1$.
\end{lemma}

\begin{proof}
Continue $\hat\cA$ along the whole component. It is finite, and
$P(s)$ stays in the open lower half-plane by
\eqref{eq:Qneg}. Hence a continuous branch of $\arg P$ takes values in
$(-\pi,0)$ and $\Delta\arg P<\pi$, even if the continued values do not close.
Also $\Delta\arg\hat\cA<0$ by \eqref{eq:Qneg}. By \eqref{eq:lambdabdy},
\begin{equation*}
  \pi d_i=\Delta\arg(d\hat\cA/ds)
  =\Delta\arg P+\Delta\arg\hat\cA<\pi .
\end{equation*}
\end{proof}

\section{Disk topology without monodromy}
\label{sec:disk}
\label{sec:theorem}

We now combine the boundary bounds of \cref{sec:boundary} with the count of zeros on the
double.

\begin{theorem}[Only the disk]
\label{thm:onlydisk}
Let $\Sigma$ have genus $g$ and $\nu\ge1$ boundary components, with five-brane
poles only on $\partial\Sigma$, no interior punctures, and single-valued meromorphic
differentials $d\cA_\pm$ satisfying the boundary reflection condition \eqref{eq:bdycond} of DGU~II.
If the solution is physically regular in the sense of \cref{sec:global}, then
\begin{equation}
  2g+\nu<2,
  \qquad\text{that is}\qquad
  (g,\nu)=(0,1):\ \Sigma\ \text{is a disk} .
\end{equation}
\end{theorem}

\begin{proof}
To recover the count of DGU~II \cite[\S6.4, (6.23)]{DHoker:2017mds},
consider the meromorphic one-form $d\cA_+$ on the double $\hat\Sigma$.
Its only poles are the $L$ simple poles on the gluing boundary.
Reflection identifies
the zeros of $d\cA_+$ in the mirror interior with those of $d\cA_-$ in
the physical interior, where $\kappa^2>0$ excludes them. On a regular
boundary arc a zero would be common to both differentials, which Hopf's
lemma excludes as in \cref{sec:identities}. The total zero order on
$\hat\Sigma$ is therefore exactly $N$, the number of zeros of $d\cA_+$ in the interior of
$\Sigma$. Since $d\cA_+$ is a meromorphic
one-form on the double, its total zero order minus its total pole order is
the degree of the canonical bundle, $\deg K_{\hat\Sigma}=2\hat g-2$, with
$\hat g=2g+\nu-1$. Hence
\begin{equation}
  N-L=2\hat g-2,\qquad N=L+4g+2\nu-4 .
  \label{eq:unpunctured-count}
\end{equation}

We now show that the boundary estimates imply $N<L$. By \cref{lem:degrees},
$N=\sum_i d_i$ and each $d_i$ is a positive integer. A pole-free boundary component
would have $d_i<1$ by \cref{lem:polefree-bound}, so every component carries poles.
Since $\cG=0$ on every boundary component and $\cG>0$ inside,
\cref{lem:rotation} gives $d_i<L_i$ on each. Summing yields
\begin{equation*}
  N=\sum_i d_i<\sum_i L_i=L .
\end{equation*}
Substitution in \eqref{eq:unpunctured-count} gives $2g+\nu<2$, which
permits only $(g,\nu)=(0,1)$.
\end{proof}

On the disk, $d_1=L-2>0$ recovers the bound $L\ge3$ of DGU~II
\cite[\S3.3]{DHoker:2017mds}. In the brane web, $L=2$ would mean two external stacks
with opposite charges by charge conservation. They would lie along a single
straight line and meet in no junction.

The annulus studied in DGU~II \cite[\S5.7]{DHoker:2017mds} shows the two ways in which
the proof excludes a surface. Suppose first
that all poles lie on one boundary component. On the other component,
\cref{lem:polefree-bound} gives $d_i<1$, whereas $d_i$ is a positive integer.
If both components carry poles, the rotation bound gives
$d_1+d_2<L_1+L_2=L$, while the global count \eqref{eq:unpunctured-count}
requires $d_1+d_2=L$.

\section{Monodromy and orientifold punctures}
\label{sec:punctures}

The boundary rotation bounds of \cref{sec:rotation} still hold in the presence
of punctures. In \cref{sec:disk}, $\lambda$ was single-valued, so the
argument principle identified the total boundary winding $\sum_i d_i$
with the number of interior zeros of $d\cA_+$. Monodromy can mix $d\cA_+$
and $d\cA_-$, making $\lambda$ multivalued. The argument principle then no
longer gives this identification, and the $d_i$ need not be integers.
We instead count, on each $\Gamma_i$, the zeros of $D\,dw$ and the
five-brane poles at which its residue vanishes, with $D\,dw$ as in
\eqref{eq:Dintro}. These are precisely the boundary poles of the ratio
$\mathcal V=\alpha\,\partial_w\cA_+/D$ studied below. Their number $z_i$ satisfies
$z_i\le d_i$ under the hypotheses below. The degree formula on the
Schottky double fixes $\sum_i z_i$ when applied to $D\,dw$ for seven-branes
or to $D^2\,dw\otimes dw$ in the presence of orientifolds.

\subsection{The invariant differential}
\label{sec:invdiff}

Consider first mutually local seven-branes. In the common D7-brane frame of
\cref{sec:sevenbranes}, monodromy shifts both differentials by the same
one-form, so their difference is single-valued. In a general duality frame
we denote the corresponding monodromy-invariant one-form by $D\,dw$. It extends meromorphically
to the double, so its zeros and poles obey the same degree formula used in
\cref{sec:disk}. We write $D$ for the coefficient of $D\,dw$ in a local coordinate and refer
to $D\,dw$ as the invariant differential. Statements that $D$ is real use a coordinate in
which the boundary is real.

\begin{lemma}[Invariant differential]
\label{lem:invdiff}
There is a complex constant $\alpha$, with $|\alpha|=1$, such that
\begin{equation}
  D:=\alpha\,\partial_w\cA_+-\bar\alpha\,\partial_w\cA_- ,
  \label{eq:Ddef}
\end{equation}
is invariant under each seven-brane monodromy $M_j$. In the common D7-brane
frame these matrices are powers $T^{n_j^2}$ of the same generator, with
strengths $n_j^2>0$ that may vary among punctures. The common invariant
direction determines $\alpha$ up to sign when at least one such puncture is
present. On an unpunctured surface in the monodromy-free class, any unit
$\alpha$ may be chosen. Under the global and local assumptions of
\cref{sec:sevenbranes}, $D\,dw$ has the following properties.

\begin{enumerate}
\item $D\,dw$ is single-valued and meromorphic on $\Sigma$, regular at the punctures, with poles only
  at the five-brane points, where it has at worst a simple pole, of residue $2\,\mathrm{Re}(\alpha
Z_\ell)\in\mathbb{R}$. The
  residue may vanish, and $D$ is then holomorphic at that point; we write $L_0$ for the number of
  five-brane points at which it does, so that $D$ has exactly $L-L_0$ poles.
\item $D$ is \emph{real} along $\partial\Sigma$, so $D\,dw$ extends to a meromorphic one-form
  $\hat D$ on the Schottky double $\hat\Sigma$.
\item $D$ has no zero in the interior of $\Sigma$, nor in its mirror.
\item Define $\mathcal{V}:=\alpha\,\partial_w\cA_+/D$. At regular points
  where $D\ne0$, we have
  \begin{equation}
    \kappa^2=|D|^2\big(1-2\,\mathrm{Re}\,\mathcal{V}\big) ,
    \qquad
    \lambda=\frac{\bar\alpha}{\alpha}\cdot\frac{\mathcal{V}}{\mathcal{V}-1} ,
    \qquad\text{and on }\partial\Sigma\quad
    \mathrm{Re}\,\mathcal{V}=\tfrac12 .
    \label{eq:nucayley}
  \end{equation}
  Thus interior positivity is equivalent to $\mathrm{Re}\,\mathcal V<\tfrac12$.
  The Cayley transform in \eqref{eq:nucayley} maps the boundary line
  $\mathrm{Re}\,\mathcal V=\tfrac12$ to the unit circle in the $\lambda$-plane.
\end{enumerate}
\end{lemma}

\begin{proof}
Invariance of $D$ means that $(\alpha,-\bar\alpha)$ is a left eigenvector of
$M_j$ with eigenvalue $1$. A non-trivial parabolic element of trace $+2$ has
$\mathsf a_j=1+i\beta_j$, with $\beta_j\in\mathbb R\setminus\{0\}$, and
$|\mathsf b_j|=|\beta_j|$ by the $SU(1,1)$ constraint. The two components of the
eigenvector equation are conjugate to one another and reduce to
\begin{equation}
  \bar{\mathsf{b}}_j=i\beta_j\,\alpha/\bar\alpha .
  \label{eq:alphacond}
\end{equation}
Since $|\bar{\mathsf b}_j/(i\beta_j)|=1$, we can choose $\alpha=e^{i\vartheta}$ with
$e^{2i\vartheta}=\bar{\mathsf b}_j/(i\beta_j)$. This determines $\alpha$ up to sign.
Commuting parabolic elements share their invariant direction, so the same $\alpha$
works for every puncture.
Changing its sign changes $D$ by a sign but leaves $|D|^2$ and $\mathcal{V}$ unchanged.

(1) The absence of independent duality holonomy makes $D$ single-valued.
In the D7-brane frame, the logarithmic terms in \eqref{eq:sevenbrane} cancel,
leaving $D=\alpha(\partial_w\cA^{\mathrm{seed}}_+-\partial_w\cA^{\mathrm{seed}}_-)$, regular at
punctures. At a five-brane pole, \eqref{eq:residues} gives the residue
$2\operatorname{Re}(\alpha Z_\ell)$; if it vanishes, the pole is removable.

(2) On $\partial\Sigma$ the
reflection condition \eqref{eq:bdycond} gives
$D=\alpha\,\partial_w\cA_++\overline{\alpha\,\partial_w\cA_+}=2\,\mathrm{Re}(\alpha\,\partial_w\cA_+)$,
which is real. Schwarz reflection therefore extends its coefficient to the mirror
by $D(\bar w)=\overline{D(w)}$, giving the one-form on the double.

(3) If $D=0$ then
$|\partial_w\cA_+|=|\partial_w\cA_-|$, because $|\alpha|=|\bar\alpha|$, hence $\kappa^2=0$, which
interior positivity excludes away from the punctures. At a seven-brane
puncture $w_j$, we assume the logarithmic divergence of $\kappa^2$ specified
in \cref{sec:sevenbranes}. Its coefficient in \eqref{eq:sevenkappa} is
proportional to $n_j^2|D(w_j)|^2$, so this source condition requires
$D(w_j)\ne0$.
Reflection gives the same conclusion on the mirror.

(4) Substituting $\partial_w\cA_+=\alpha^{-1}\mathcal V D$ and
$\partial_w\cA_-=\bar\alpha^{-1}(\mathcal V-1)D$ into the definitions of
$\lambda$ and $\kappa^2$ gives \eqref{eq:nucayley}.
\end{proof}

Five-brane points at which the residue of $D$ vanishes must be retained in the count: in the D7-brane frame the
residue of $D$ is proportional to the NS5 charge, so a pure D5 pole is regular
for $D$ but is a pole of $\mathcal V$ \cite[below (3.71)]{DHoker:2017zwj}.

\subsection{Counting the boundary poles}
\label{sec:count}

We count the boundary poles of $\mathcal V$ using the degree of $\hat D$,
that is, its total zero order minus its total pole order.

\begin{lemma}[Boundary pole count]
\label{lem:dcount}
Let $z_i$ denote the number of poles of $\mathcal V$ on $\Gamma_i$.
On the regular boundary arcs these poles occur precisely at the zeros of $D$.
At a five-brane pole, $\mathcal V$ has a pole precisely when the residue of
$D\,dw$ vanishes. In that case $D$ is finite and nonzero.
All poles of $\mathcal V$ on $\Gamma_i$ are simple. Summing both contributions
over the boundary components gives
\begin{equation}
  \begin{aligned}
  \sum_{i=1}^\nu z_i
    &=\underbrace{(L-L_0)+2\hat g-2}_{\text{boundary zeros of }D\,dw}
      +\underbrace{L_0}_{\substack{\text{five-brane poles with}\\
                                    \operatorname{Res}_{p_\ell}(D\,dw)=0}}\\
    &=L+2\hat g-2=L+4g+2\nu-4 .
  \end{aligned}
  \label{eq:dtotal}
\end{equation}
\end{lemma}

\begin{proof}
By \cref{sec:boundary}, neither of $\partial_w\cA_\pm$ vanishes on an open
boundary arc. Thus a zero of $D$ there gives a pole of $\mathcal V$ of the same
order. At a five-brane point with $\operatorname{Re}(\alpha Z_\ell)\ne0$, the
poles in numerator and denominator cancel and
$\mathcal V\to\alpha Z_\ell/[2\operatorname{Re}(\alpha Z_\ell)]$.
At the remaining $L_0$ points, the numerator still has a simple pole while
$D$ is holomorphic.

All these poles are simple, so that their number equals their total order. This follows
from \eqref{eq:lambdaprime}. Dividing \eqref{eq:Ddef} by $\alpha\,\partial_w\cA_-$ gives
\begin{equation}
  \lambda-\frac{\bar\alpha}{\alpha}=\frac{D}{\alpha\,\partial_w\cA_-} ,
  \label{eq:lambdaD}
\end{equation}
with $\partial_w\cA_-$ nonzero on every regular boundary arc. At a boundary zero $x_0$ of
$D$ of order $m$ away from the poles, \eqref{eq:lambdaD} gives
$\lambda-\bar\alpha/\alpha\sim c\,(w-x_0)^m$ and $\lambda'\sim mc\,(w-x_0)^{m-1}$, which
\eqref{eq:lambdaprime} forbids unless $m=1$. At a five-brane point $\lambda$ is still holomorphic,
with $\lambda(p_\ell)=-Z_\ell/\overline{Z_\ell}$ of unit modulus. Applying Hopf's lemma
to the extended $\log|\lambda|$ gives $\lambda'(p_\ell)\ne0$ at the pole as well.
When the residue of $D\,dw$ vanishes, \eqref{eq:lambdaD} gives
$\lambda(p_\ell)=\bar\alpha/\alpha$. Since
$\partial_w\cA_-\sim Z_-^\ell/(w-p_\ell)$, taking the limit in
$D=\alpha\,\partial_w\cA_-(\lambda-\bar\alpha/\alpha)$ gives
$D(p_\ell)=\alpha Z_-^\ell\lambda'(p_\ell)\ne0$.
Thus each of the $L_0$ points contributes precisely one simple pole of $\mathcal V$.

The same degree formula used in \cref{sec:disk} gives $L-L_0+2\hat g-2$
zeros of $\hat D$, counted with multiplicity. They all lie on the boundary
by \cref{lem:invdiff} and are simple by the preceding argument. Each boundary point is counted once on the double. Adding the
$L_0$ five-brane points with vanishing residue, which are poles of $\mathcal V$ but not of
$D$, gives \eqref{eq:dtotal}.
\end{proof}

For a disk, $g=0$ and $\nu=1$, so \eqref{eq:dtotal} gives $z_1=L-2$
for the number of poles of $\mathcal V$ on its single boundary component.
For the two-pole disks of Chaney and Uhlemann this boundary count vanishes
regardless of the number of interior seven-brane punctures
\cite[\S\S3.4--3.5]{Chaney:2018gjc}. For example, the solution denoted by $+_{N,M,k,j}$
in their notation has two poles and two punctures at distinct interior points
\cite[\S4.4, Fig.~7]{Chaney:2018gjc}.

\subsection{Orientifold punctures and the sign of \texorpdfstring{$D$}{D}}
\label{sec:O7}

An orientifold seven-plane has monodromy $M_{\mathrm{O7}^{\pm}}=-T^{\pm4}$.
More generally, $-T^k$ describes an $\mathrm{O7}^-$ with $k+4$ D7-branes or, for
$k\ge4$, an $\mathrm{O7}^+$ with $k-4$ D7-branes \cite[\S II.1]{Uhlemann:2019lge}.
We restrict to $k=n_j^2\ge0$, which excludes an $\mathrm{O7}^-$ with fewer than four D7-branes. For $k>0$, this monodromy commutes with non-trivial seven-brane
monodromies only when they lie in the same parabolic subgroup. For $k=0$, it is the
central element $-\mathbf{1}$, corresponding to an $\mathrm{O7}^-$ with four D7-branes
and no logarithmic axion monodromy.

In the common D7-brane frame, continuation around the puncture acts as
\cite[\S II.5]{Uhlemann:2019lge}
\begin{equation}
  \cA_\pm\ \longmapsto\ -\Big(\cA_\pm+\frac{ik}{2}\big(\cA_+-\cA_-\big)\Big) ,
  \label{eq:O7mono}
\end{equation}
which in the notation of \eqref{eq:su11} is $M=-\mathsf P$, with $\mathsf P$ the trace-$+2$ parabolic element of
\cref{lem:invdiff} that a seven-brane puncture of strength $k$ carries. The left eigenvector of $\mathsf P$
with eigenvalue $1$ is $(\alpha,-\alpha)$ with $\alpha$ real, so that
we choose $\alpha=1$ and write $D=\partial_w\cA_+-\partial_w\cA_-$, as in
\cite{Uhlemann:2019lge}. For $M=-\mathsf P$ the eigenvalue is $-1$:
\begin{equation}
  D\ \longmapsto\ -D .
  \label{eq:Dflip}
\end{equation}
For $k=0$, the central element $-\mathbf{1}$ imposes no constraint
on $\alpha$, but still sends $D$ to $-D$. No nonzero linear combination is fixed by this
monodromy. Thus $D$ is double-valued: its local branches differ
only by a sign, with $D\mapsto-D$ around each orientifold puncture. Equivalently, $D\,dw$
takes values in a flat $\mathbb Z_2$ bundle. The single-valuedness in
\cref{lem:invdiff}(1) fails, while $D^2\,dw\otimes dw$ remains single-valued.

Uhlemann presented closed-form solutions with one orientifold puncture, obtained as a quotient
of disk solutions with two seven-brane punctures \cite[\S II.4]{Uhlemann:2019lge}. In a local coordinate $\xi$ vanishing at the puncture, their
differentials behave as \cite[\S II.5]{Uhlemann:2019lge}
\begin{equation}
  \partial_\xi\cA_\pm=\frac1{\sqrt{-\xi}}\Big[\mathfrak{a}_\pm(\xi)+\frac{k}{4\pi}\ln(-\xi)\,
  \big(\mathfrak{a}_+(\xi)-\mathfrak{a}_-(\xi)\big)\Big] ,
  \qquad k=n_j^2\ge0 ,
  \label{eq:O7local}
\end{equation}
with $\mathfrak{a}_\pm$ holomorphic at $\xi=0$; they are written $\partial_u\cA^{(0)}_\pm$
there. This behavior was proposed in \cite{Uhlemann:2019lge} as the starting point for solutions
with several orientifold punctures. Following this proposal, we require \eqref{eq:O7local}
at every orientifold puncture, independently of the topology of $\Sigma$ and the number of
punctures. We also assume that
$\mathsf{C}:=\mathfrak{a}_+(0)-\mathfrak{a}_-(0)\ne0$. The sign change of both
differentials comes from the common factor $1/\sqrt{-\xi}$, and a constant phase in
the choice of branch is absorbed in $\mathfrak{a}_\pm$. The logarithmic terms cancel in the
difference, so
\begin{equation}
  D=\partial_\xi\big(\cA_+-\cA_-\big)
   =\frac{\mathfrak{a}_+-\mathfrak{a}_-}{\sqrt{-\xi}}
   \;\xrightarrow[\ \xi\to0\ ]{}\;\frac{\mathsf{C}}{\sqrt{-\xi}} .
  \label{eq:Cdef}
\end{equation}
We say that $D$ has a \emph{half-order pole}. The change $\sqrt{-\xi}\to-\sqrt{-\xi}$ on
a circuit is the sign change \eqref{eq:Dflip}. Expanding the difference of
squares in \eqref{eq:kappaG} gives
\begin{equation}
  \kappa^2=\frac1{|\xi|}\Big[\,|\mathfrak{a}_-|^2-|\mathfrak{a}_+|^2+\frac{k}{2\pi}\ln\frac1{|\xi|}\,\big|\mathfrak{a}_+-\mathfrak{a}_-\big|^2\Big] .
  \label{eq:O7kappa}
\end{equation}
If $k>0$ the last term dominates and
\begin{equation}
  \kappa^2=\frac{k|\mathsf{C}|^2}{2\pi}\,\frac{\ln(1/|\xi|)}{|\xi|}+\ldots ,
  \label{eq:O7div}
\end{equation}
which is positive and divergent, as in the seven-brane case \eqref{eq:sevenkappa}.
For $k=0$ the logarithm is absent, and
$\kappa^2=\big(|\mathfrak{a}_-(0)|^2-|\mathfrak{a}_+(0)|^2\big)/|\xi|+\mathcal{O}(1)$,
which is positive when $|\mathfrak a_-(0)|>|\mathfrak a_+(0)|$.
For $k<0$ the logarithmic term is negative and divergent, so the restriction to $k\ge0$ made
at the beginning of this subsection is required by regularity, as noted in
\cite[\S II.1]{Uhlemann:2019lge}. Likewise, \eqref{eq:sevenkappa} is negative near a
seven-brane puncture with $n_j^2<0$ and $D(w_j)\ne0$.

The condition $\mathsf C\ne0$ is an assumption of the orientifold theorem (\cref{thm:extended}).
If $\mathsf C=0$, then $\mathfrak a_+(0)=\mathfrak a_-(0)$, and there are two cases.
If the common value is nonzero, regularity already fails locally: either
$\mathfrak a_+=\mathfrak a_-$ identically and $\kappa^2$
vanishes, or $\mathfrak a_--\mathfrak a_+=b\,\xi^m+\ldots$ with $b\ne0$ and $m\ge1$. In the
second case the bracket in \eqref{eq:O7kappa} is
$2\,\mathrm{Re}\big(\overline{\mathfrak a_+(0)}\,b\,\xi^m\big)+\ldots$, which is negative
for some directions of $\xi$, however small $|\xi|$ is.
The assumption therefore excludes a possibly regular solution only if
$\mathfrak a_+(0)=\mathfrak a_-(0)=0$. In that case $D$ vanishes at the puncture and has no
half-order pole, so the count of \cref{sec:extended} would change. We do not consider this
case.

\subsection{Boundary winding in the presence of monodromy}
\label{sec:noninteger}

In the common D7-brane frame, $\alpha=1$ and $D=\partial_w\cA_+-\partial_w\cA_-$. By
\cref{sec:sevenbranes}, continuation counterclockwise around a seven-brane puncture $w_j$ sends
$\partial_w\cA_\pm\mapsto\partial_w\cA_\pm+\frac{in_j^2}{2}D$, and by \eqref{eq:O7mono} an
orientifold puncture adds an overall minus sign. In both cases the ratio
$\mathcal{V}=\partial_w\cA_+/D$ changes by
\begin{equation}
  \mathcal{V}\ \longmapsto\ \mathcal{V}+\frac{in_j^2}{2} .
  \label{eq:nushift}
\end{equation}
The sign change cancels in $\mathcal{V}$, and likewise in $\lambda$. The imaginary shift
remains and can make $d_i$ non-integral. At an orientifold puncture, \eqref{eq:O7local} gives
$\mathcal{V}$ the same logarithm $\frac{n_j^2}{4\pi}\ln(-\xi)$ as $f$ has at a seven-brane
puncture. The sum over $j$ in \eqref{eq:sevenbrane} therefore runs over the punctures of both
types, and an orientifold puncture contributes with its $n_j^2$ in the same way as a seven-brane
puncture. The next lemma gives the shift accumulated along a boundary component.

\begin{lemma}[Shift along a boundary component]
\label{lem:shift}
Assume the global construction of \cref{sec:sevenbranes}.
Continued once around $\Gamma_i$ in the positive direction, $\mathcal{V}$ returns to
$\mathcal{V}+is_i$. Let $\omega_i$ be the harmonic measure of $\Gamma_i$ in $\Sigma$ --- the harmonic function on
$\Sigma$ with boundary value $1$ on $\Gamma_i$ and $0$ on the other components, so that
$\sum_i\omega_i\equiv1$. Then
\begin{equation}
  s_i=\frac12\sum_j n_j^2\,\omega_i(w_j)\;\ge\;0 ,
  \qquad
  \sum_i s_i=\frac12\sum_j n_j^2 .
  \label{eq:shift}
\end{equation}
The coefficients $n_j^2$ are non-negative for both types of puncture. Since harmonic
measure is strictly positive in the interior, $s_i>0$ whenever at least one puncture has
positive strength.
\end{lemma}

\begin{proof}
In the D7-brane frame the $f$-terms in $\partial_w\cA_\pm$ are identical and cancel in $D$, so
$D=\partial_w\cA^{\mathrm{seed}}_+-\partial_w\cA^{\mathrm{seed}}_-$ and
\begin{equation*}
  \mathcal{V}=f+\frac{\partial_w\cA^{\mathrm{seed}}_+}{D} .
\end{equation*}
By the global assumption of \cref{sec:sevenbranes} the second term is single-valued: the seed is
single-valued, or, in the orientifold case, changes sign together with $D$. The change of
$\mathcal{V}$ around $\Gamma_i$ is therefore that of $f$. Since $\mathrm{Re}\,f$ is
single-valued, only $\mathrm{Im}\,f$ changes, and the change of $f$ is $is_i$ with
$s_i=\oint_{\Gamma_i}d\,\mathrm{Im}\,f$. By the
Cauchy--Riemann equations, $d\,\mathrm{Im}\,f=-\partial_n\mathrm{Re}\,f\,d\ell$ along the
boundary, with $\partial_n$ the inward normal derivative, and \eqref{eq:sevenbrane} gives
\begin{equation*}
  s_i=\frac1{4\pi}\sum_j n_j^2\oint_{\Gamma_i}\partial_n\mathsf{g}(\cdot,w_j)\,d\ell
    =\frac12\sum_j n_j^2\,\omega_i(w_j) ,
\end{equation*}
using the standard relation
$\frac1{2\pi}\oint_{\Gamma_i}\partial_n\mathsf{g}(\cdot,w_j)\,d\ell=\omega_i(w_j)$ between the
Green's function and harmonic measure.
\end{proof}

The total $\sum_i s_i=\frac12\sum_j n_j^2$ is the sum of the local shifts \eqref{eq:nushift},
as it must be: $\mathcal V$ is multi-valued, but $d\mathcal{V}$ is a single-valued holomorphic
one-form in the punctured interior, and the sum of the boundary components, pushed slightly
into the interior, is homologous there to the sum of small circles around the punctures. The global
assumption enters only in the distribution of this total among the boundary components.
Without it, $\mathrm{Im}\,(\mathcal{V}-f)$ could have a nonzero period along each $\Gamma_i$,
which would add to $s_i$ in \eqref{eq:shift}. These periods sum to zero but need
not have a fixed sign, so $s_i\ge0$ would no longer follow. This limitation
is noted in \cref{sec:discussion}.

\begin{lemma}[The fractional part]
\label{lem:lift}
Let $z_i$ be as in \cref{lem:dcount}, and let $t=\mathrm{Im}\,\mathcal{V}$. For a base point at which $\mathcal{V}$
is finite, with $t_0$ the value of $t$ there,
\begin{equation}
  d_i=z_i+\varepsilon_i ,
  \qquad
  \varepsilon_i=\frac1\pi\Big[\arctan\big(2(t_0+s_i)\big)-\arctan(2t_0)\Big]\in[0,1) ,
  \label{eq:lift}
\end{equation}
and for a base point at a pole of $\mathcal{V}$, one has $d_i=z_i$. A five-brane point is of this type
only if the residue of $D$ vanishes there. Otherwise $\mathcal{V}$ is finite and
$t_0=\operatorname{Im}(\alpha Z_\ell)/[2\operatorname{Re}(\alpha Z_\ell)]$.
In all cases $d_i\ge z_i$. At a finite base point equality holds precisely when $s_i=0$,
equivalently when every puncture has zero strength.
\end{lemma}

\begin{proof}
By \eqref{eq:nucayley} and $\mathrm{Re}\,\mathcal{V}=\tfrac12$ on $\partial\Sigma$,
$\arg\lambda=2\arctan(2t)-\pi+\arg(\bar\alpha/\alpha)$, so $\arg\lambda$ is a function of $t$ alone
and $\pi d_i=\Delta_{\Gamma_i}\Theta$. Here $\Theta$ is obtained from $\arctan(2t)$ by
adding multiples of $\pi$ so that it remains continuous when $t$ passes through infinity.
By \eqref{eq:psimono}, $t$ is strictly increasing between its poles, and it passes through $\infty$ exactly at the poles
of $\mathcal{V}$, all of which are simple by \cref{lem:dcount}; each passage advances $\Theta$ by $\pi$,
contributing $\pi z_i$. The endpoints differ by the monodromy, $t_{\rm end}=t_0+s_i$, contributing
the bracket, which lies in $[0,\pi)$ because $s_i\ge0$ by \cref{lem:shift} and vanishes exactly when
$s_i=0$. For a base point at a pole of $\mathcal{V}$ the circuit runs from $t=-\infty$ to $t=+\infty$, so
$z_i-1$ passages occur inside it and the endpoints contribute $\pi$: again $\pi z_i$, the monodromy
being invisible between the two ends of the real line.
\end{proof}

\begin{lemma}[No pole-free component]
\label{lem:polefree}
Assume the global construction of \cref{sec:sevenbranes}, with $D$ single-valued up to a sign.
In a regular solution every boundary component carries a five-brane pole: $L_i\ge1$ for all $i$.
\end{lemma}

\begin{proof}
Suppose $L_i=0$. By \cref{lem:polefree-bound}, $d_i<1$. Since $z_i$ is a non-negative integer
and $z_i\le d_i$ by \cref{lem:lift}, it vanishes. Hence $D$, which is real on $\Gamma_i$, has
no zero there by \cref{lem:dcount}. It cannot change sign on a circuit of $\Gamma_i$, so it returns to itself, and
the integral of $D\,dw$ along $\Gamma_i$ is real and nonzero.

We show that this integral must be imaginary. Continuation around $\Gamma_i$ acts as
$\cA_\pm\mapsto(M\cA)_\pm+a_\pm$, with $M$ the monodromy and $a_\pm$ constants. The linear
part leaves $\cG$ invariant. By \eqref{eq:kappaG} and \eqref{eq:Bdef}, the constants $a_\pm$
change $\cG$ by
\begin{equation*}
  2\,\mathrm{Re}\Big[\big(\bar a_+-a_-\big)\big(\cA_++\overline{\cA_-}\big)\Big]+\text{const} .
\end{equation*}
If $a_-\ne\bar a_+$ this is not constant, since otherwise $\lambda$ would be a constant of
unit modulus and $\kappa^2$ would vanish. Since $\cG$ is single-valued, $a_-=\bar a_+$.
Since $D$ returns to itself, $M$ leaves $\alpha\cA_+-\bar\alpha\cA_-$ invariant, so the integral of
$D\,dw=d(\alpha\cA_+-\bar\alpha\cA_-)$ along $\Gamma_i$ is
$\alpha a_+-\bar\alpha a_-=2i\,\mathrm{Im}(\alpha a_+)$. This contradicts the first paragraph.
\end{proof}

\subsection{Topology constraint with seven-branes}
\label{sec:sevenbrane-theorem}

We combine the count of \cref{lem:dcount} with the bounds of \cref{sec:rotation} and
\cref{lem:lift}.

\begin{theorem}[Only the disk, with seven-branes]
\label{thm:sevenbranes}
Let $\Sigma$ have genus $g$ and $\nu\ge1$ boundary components, with five-brane poles only
on $\partial\Sigma$, and any number of punctures carrying mutually commuting parabolic monodromies
of trace $+2$. Assume the global construction \eqref{eq:sevenbrane} of
\cref{sec:sevenbranes}, with no independent duality holonomy, and $D(w_j)\ne0$ at each
puncture. If the solution is physically regular in the sense
of \cref{sec:global}, then
\begin{equation}
  2g+\nu<2 ,
  \qquad\text{that is}\qquad
  (g,\nu)=(0,1):\ \Sigma\ \text{is a disk} .
\end{equation}
\end{theorem}

\begin{proof}
The function $\cG$ vanishes on $\partial\Sigma$ and is positive in the interior, including at
the punctures, so the hypothesis of \cref{lem:rotation} holds on every component.
By \cref{lem:polefree} every component carries a pole. Take the base point of $d_i$ at a
five-brane pole. Then
\cref{lem:rotation} gives $d_i<L_i$ and \cref{lem:lift} gives $z_i\le d_i$. Since $z_i$ is an
integer, $z_i\le L_i-1$. Summing over the $\nu$ components and using \eqref{eq:dtotal},
\begin{equation*}
  L+4g+2\nu-4=\sum_i z_i\le L-\nu .
\end{equation*}
Thus $4g+3\nu\le4$, and $\Sigma$ is a disk.
\end{proof}

Without punctures $s_i=0$ and $d_i=z_i$, so this proof reduces to that of \cref{thm:onlydisk},
with the boundary poles of $\mathcal V$ counted in place of the interior zeros of $\lambda$.

\subsection{Topology constraint with orientifolds}
\label{sec:extended}

With orientifolds $D$ is defined only up to a sign, and we count with $D^2$ instead.

\begin{lemma}[The count with orientifolds]
\label{lem:d2count}
Assume \eqref{eq:O7local} with $\mathsf C\ne0$ at each orientifold puncture. Let $I_-$ be the
number of such punctures, and let $z_i$ be as in \cref{lem:dcount}. Then
\begin{equation}
  \sum_{i=1}^\nu z_i=L+I_-+4g+2\nu-4 .
  \label{eq:d2total}
\end{equation}
\end{lemma}

\begin{proof}
$D$ is defined only up to sign, but $Q:=D^2\,dw\otimes dw$ is single-valued and meromorphic, and real on
$\partial\Sigma$ because $D$ is, so $Q$ doubles to a meromorphic quadratic differential $\hat Q$ on
$\hat\Sigma$. A quadratic differential has twice the degree of a one-form, so its
number of zeros minus its number of poles, counted with orders, is $4\hat g-4$.
Its poles are double at each of the $L-L_0$
five-brane points of non-vanishing residue, which lie on the fixed locus, and simple at each
orientifold puncture and at its mirror image by \eqref{eq:Cdef}; a seven-brane puncture contributes none,
and neither does one of the $L_0$ five-brane points at which $D$ is holomorphic. Its zeros: none in
the interior or its mirror, since $D=0$ forces $\kappa^2=0$; and each zero of $D$ on $\partial\Sigma$
is simple by \cref{lem:dcount}, hence a double zero of $Q$. So
\begin{equation*}
\underbrace{2\big(\sum_i z_i-L_0\big)}_{\text{zero orders}}
-\underbrace{2(L-L_0)}_{\text{boundary pole orders}}
-\underbrace{2I_-}_{\text{punctures and mirrors}}
=4\hat g-4 ,
\end{equation*}
which gives \eqref{eq:d2total}.
\end{proof}

\begin{theorem}[Only the disk, with orientifolds]
\label{thm:extended}
Let $\Sigma$ have genus $g$ and $\nu\ge1$ boundary components, with five-brane
poles only on $\partial\Sigma$. Allow puncture monodromies
$M_j=\sigma_jT^{k_j}$ in a common D7-brane frame, where $\sigma_j=\pm1$ and $k_j=n_j^2\ge0$.
This includes the central monodromy $-\mathbf{1}$ when $\sigma_j=-1$ and $k_j=0$.
Assume the global construction \eqref{eq:sevenbrane} of \cref{sec:sevenbranes}, with $D$
single-valued up to a sign, $D\mapsto-D$ around each orientifold puncture, and no independent
duality holonomy. Assume $D(w_j)\ne0$ at each seven-brane puncture, and \eqref{eq:O7local}
with $\mathsf{C}\ne0$ at each orientifold puncture. If the solution is physically regular in the
sense of \cref{sec:global}, then
\begin{equation}
  (g,\nu)=(0,1)
  \qquad\text{and}\qquad
  I_-\le1 :
  \quad \Sigma\ \text{is a disk with at most one orientifold puncture} .
  \label{eq:extended}
\end{equation}
\end{theorem}

\begin{proof}
\Cref{lem:shift,lem:lift,lem:polefree} hold for punctures of either type, so
$z_i\le L_i-1$ on every component, as in the proof of \cref{thm:sevenbranes}.
The count \eqref{eq:d2total} gives
\begin{equation*}
  L+I_-+4g+2\nu-4=\sum_i z_i\le L-\nu ,
  \qquad\text{that is}\qquad
  I_-+4g+3\nu-4\le0 .
\end{equation*}
Since $\nu\ge1$, this requires $(g,\nu)=(0,1)$ and $I_-\le1$.
\end{proof}

The counts alone would allow solutions without five-brane poles. For $L=0$,
every $z_i$ vanishes by \cref{lem:polefree-bound,lem:lift}, and
\eqref{eq:d2total} gives $I_-+4g+2\nu-4=0$, satisfied by an annulus without orientifolds or
a disk with two orientifolds. Both are excluded by \cref{lem:polefree}.

The fractional winding is realized by the regular one-pole disk solutions of
\cite[\S III.2]{Uhlemann:2019lge}, denoted $\mathsf I^\pm_{N,j}$ there. Their $N$ is a
five-brane charge, and their $j$ is the exponent $k=n_1^2$ of the monodromy $-T^k$.
With one orientifold, \eqref{eq:d2total} gives $z_1=0$, so
$0<d_1=\varepsilon_1<1$ for $k>0$. On the disk \cref{lem:shift} gives $s_1=k/2$. The residue
$Z_+$ of these solutions is real, so $t_0=0$ for a base point at the five-brane pole, and
\eqref{eq:lift} gives
\begin{equation*}
  d_1=\tfrac1\pi\arctan k ,
\end{equation*}
which lies between $0$ and $\tfrac12$ and is smaller than $L_1=1$.
One five-brane pole thus suffices in the presence of an orientifold, whereas
\cref{thm:onlydisk} requires at least three without punctures and
\eqref{eq:dtotal} requires at least two with seven-branes alone.

\section{The external legs of the web}
\label{sec:consequences}
\label{sec:legs}

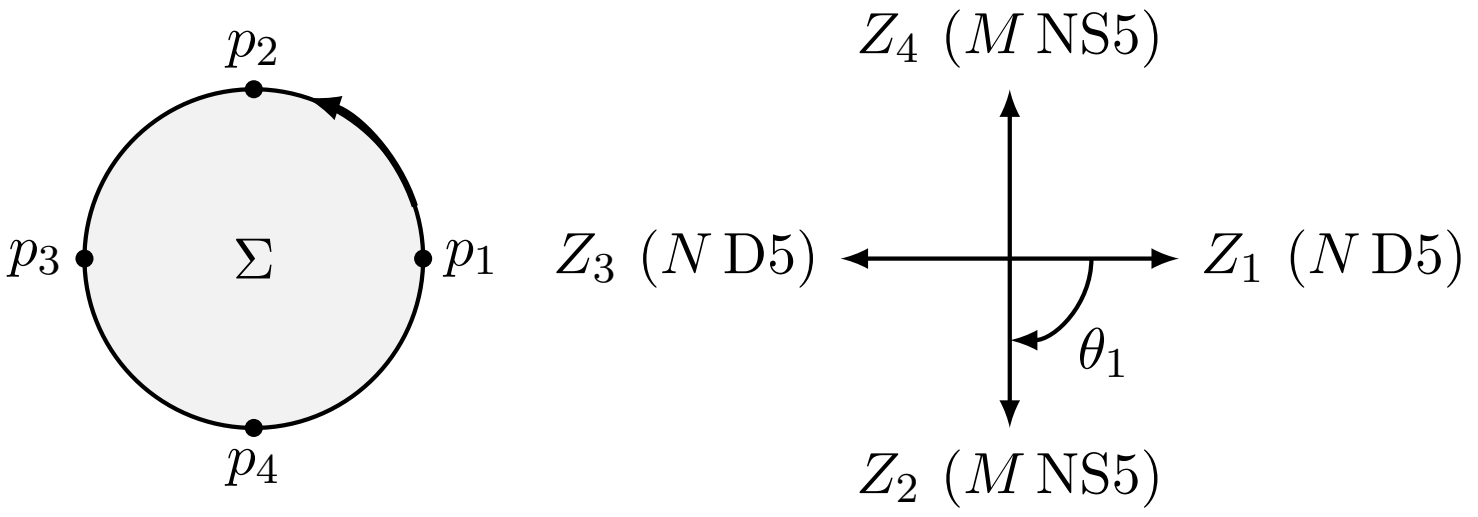
\begin{figure}[htbp]
\centering
\begin{tikzpicture}[>=latex,thick]
  \begin{scope}
    \fill[gray!10] (0,0) circle (1.12);
    \draw (0,0) circle (1.12);
    \node at (0,0) {$\Sigma$};
    \foreach \angle/\idx/\pos in {0/1/right,90/2/above,180/3/left,270/4/below}{
      \fill (\angle:1.12) circle (1.7pt) node[\pos] {$p_{\idx}$};
    }
    \draw[->,very thick] (18:1.12) arc (18:72:1.12);
  \end{scope}
  \begin{scope}[xshift=5cm]
    \foreach \angle/\idx/\pos/\stack in
      {0/1/right/{N\,\mathrm{D5}},-90/2/below/{M\,\mathrm{NS5}},
       180/3/left/{N\,\mathrm{D5}},90/4/above/{M\,\mathrm{NS5}}}{
      \draw[->] (0,0) -- (\angle:1.12) node[\pos] {$Z_{\idx}$ ($\stack$)};
    }
    \draw[->] (0:0.54) arc (0:-90:0.54);
    \node at (-45:0.88) {$\theta_1$};
  \end{scope}
\end{tikzpicture}
\caption{Charge ordering for the $+_{N,M}$ web. Left: positive traversal of
the disk boundary takes $p_1$ to $p_2$ counterclockwise. The angular
spacing of the poles depends on the choice of disk coordinate. Right: the
charge directions are drawn with unit length and rotated together to place
the D5 stacks horizontally. They turn clockwise
from $Z_1$ to $Z_2$. For $+_{N,M}$ each step has $\theta_\ell=-\pi/2$.
In general the angles may differ, but their sum is $-2\pi$.}
\label{fig:charge-order}
\end{figure}

Fluder and Uhlemann \cite[\S III.1]{Fluder:2020pym} observed that
regularity enforces agreement between the order of the boundary poles and the order
of the external five-branes around the junction. We recover this ordering from \cref{lem:rotation}.
On an unpunctured disk, label the poles in positive cyclic order along the boundary, with indices
understood modulo $L$, and let $I_\ell$ be the open boundary arc from $p_\ell$
to $p_{\ell+1}$ in this orientation. Write $\theta_\ell=\Arg(Z_{\ell+1}/Z_\ell)\in(-\pi,\pi]$, with $\Arg$ the principal value, for the
rotation of the residue from one pole to the next along the boundary. The direction of $Z_\ell$
in the complex plane represents the direction of the five-brane charge. Figure~\ref{fig:charge-order} illustrates
the two orientations for the four-stack web discussed below.

\begin{corollary}[External charge ordering]
\label{cor:legs}
For a regular disk solution with $L$ poles and no punctures,
\begin{equation}
  \theta_\ell\in(-\pi,0)\ \ \text{for every }\ell ,
  \qquad
  \sum_{\ell=1}^{L}\theta_\ell=-2\pi .
  \label{eq:legs}
\end{equation}
The five-brane charge directions follow the cyclic order of the poles and make one complete
clockwise turn.
\end{corollary}

\begin{proof}
The disk has a single boundary component, on which $\cG$ takes its minimum, so
\cref{lem:rotation} gives $\delta_{I_\ell}<\pi$, while
\eqref{eq:lambdabdy} and \eqref{eq:psimono} give
$\delta_{I_\ell}=\tfrac12\Delta_{I_\ell}\arg\lambda>0$.
Recall that $\delta_{I_\ell}$ is the change of $\arg\partial_w\cA_+$ along $I_\ell$ in an
upper-half-plane coordinate $w=x+iy$. Near each end of $I_\ell$ the differential is
dominated by the pole there. Just after $p_\ell$ we have
$\partial_w\cA_+\simeq Z_\ell/(x-p_\ell)$ with $x-p_\ell>0$, so the phase starts at
$\arg Z_\ell$. Just before $p_{\ell+1}$ we have
$\partial_w\cA_+\simeq Z_{\ell+1}/(x-p_{\ell+1})$ with $x-p_{\ell+1}<0$, so the phase ends at
$\arg Z_{\ell+1}+\pi$. Hence $\delta_{I_\ell}\equiv\theta_\ell+\pi\pmod{2\pi}$. Since
$\theta_\ell+\pi\in(0,2\pi]$ and $\delta_{I_\ell}\in(0,\pi)$, the two are equal, and
$\theta_\ell\in(-\pi,0)$. Finally, $\sum_\ell\delta_{I_\ell}=\pi d_1$ by \cref{sec:rotation},
and $d_1=N=L-2$ by \cref{lem:degrees} and \eqref{eq:unpunctured-count}. Therefore
$\sum_\ell\theta_\ell=\pi(L-2)-\pi L=-2\pi$.
\end{proof}

Since the charge directions turn strictly clockwise through a total of $2\pi$,
distinct boundary poles cannot carry five-brane charges in the same direction, and
consecutive legs cannot be anti-parallel.
A single pole may carry the total charge of several coincident five-branes.
Non-consecutive legs may be anti-parallel, as in the $+_{N,M}$ web of
\cref{fig:charge-order} \cite[\S2.2]{Bergman:2018hin}, where each anti-parallel pair is
separated by one other stack in the cyclic order.

As counted in \cref{sec:global}, a regular disk solution with prescribed charges $Z_\ell$
amounts to a solution of \eqref{eq:regcond} for the cross-ratios of the poles and the constant
$\cA^0$, and such solutions are generically isolated. In addition $\kappa^2>0$ requires the $L-2$ zeros of
$\partial_w\cA_+=\sum_\ell Z_\ell/(w-p_\ell)$, the points $s_n$ of \eqref{eq:zeroparam},
to lie in the interior. The cyclic ordering of
\cref{cor:legs} is necessary for such a solution. Whether it is sufficient, and whether
the solution is unique, remains open. The equations have been solved numerically for up to
ten poles in \cite{Fluder:2020pym}, which also reports roots that violate $\kappa^2>0$.

\section{Discussion}
\label{sec:discussion}

In the monodromy-free class of DGU~II, a regular solution with boundary
five-brane poles requires $\Sigma$ to be a disk. This excludes the unpunctured
annulus of DGU~II \cite[\S5.7]{DHoker:2017mds} as well as higher genus,
even for a surface with a single boundary component. This proves, within the
DGU~II class, the disk restriction anticipated by Kaidi and Uhlemann from
the M-theory description of five-brane webs \cite[\S3]{Kaidi:2018zkx}. For webs with a regular
supergravity dual in this class, \cref{thm:onlydisk} justifies the choice of a disk
as the base of the $S^2$ fibration in the symmetry TFT derivation of
Apruzzi et al.\ \cite[\S5.1]{Apruzzi:SymTFT}.

For mutually local seven-branes,
\cref{thm:sevenbranes} excludes both positive genus and multiple boundary
components under the global and local assumptions of \cref{sec:sevenbranes}.
In particular, no choice of parameters in the annulus ansatz of Lozano et al.\
\cite[\S3]{Lozano:AdS6Tduals} yields a regular solution with standard localized
D7-brane sources.
Allowing an overall minus sign around orientifolds in the global construction,
and imposing the non-degenerate local asymptotics of \cref{sec:O7}, we find
that $\Sigma$ is again a disk, with at most one orientifold puncture. In both cases every boundary
component carries a five-brane pole (\cref{lem:polefree}), so seven-branes and orientifolds
alone cannot source a regular solution. More general puncture
asymptotics and independent duality holonomy remain outside this analysis.

The restriction to localized sources matters for the topology question.
The explicit $\ads_6$ annulus of Lozano et al.\ has smeared sources on both
boundary components \cite[\S5]{Lozano:AdS6Tduals}. A continuous source
distribution along a boundary precludes the decomposition into smooth arcs
between isolated poles used to bound the phase rotation. Whether smeared
sources admit surfaces of positive genus or with more than two boundary
components remains open. A physical role for smearing also arises in the
M-theory $\ads_5$ duals proposed by Bah et al.\ \cite{Bah:ADduals,Bah:M5Sources}
and generalized by Couzens et al.\ \cite{Couzens:IrregularSphere}, where a smeared
M5-brane source realizes the irregular puncture of a four-dimensional
Argyres--Douglas theory.

\Cref{thm:sevenbranes,thm:extended} rely on a charge direction preserved by every
puncture monodromy. Such a direction exists for the parabolic monodromies of mutually
local seven-branes and, up to a sign, for orientifolds. The latter include the $D_4$-type
puncture with monodromy $-\mathbf 1$ produced by a $\mathbb Z_2$ quotient
\cite[\S2.1]{Apruzzi:Sfolds}, which is the $k=0$ orientifold of \cref{sec:O7}
\cite{Uhlemann:2019lge}. Combinations of mutually non-local seven-branes, which realize
exceptional flavor symmetries in five-brane webs \cite{DeWolfe:1999hj}, need not have
such a direction. They can enter in two ways. First, a single puncture can carry their
combined monodromy. In the S-fold solutions of Apruzzi et al.\ \cite[\S3]{Apruzzi:Sfolds},
a puncture of type $E_{6,7,8}$ or $H_{0,1,2}$ sits at the center of a disk. Its monodromy
is elliptic and has no eigenvalue $\pm1$, so it preserves no charge direction, even up to
a sign. Second, distinct punctures can carry non-commuting monodromies. The
construction of such solutions was raised as an open problem in DGU~III
\cite[\S6]{DHoker:2017zwj}. In both cases the invariant differential $D\,dw$ is
unavailable, and its zero and pole count on the double cannot be applied. Our argument
therefore does not determine whether regularity still forces $\Sigma$ to be a disk in
this broader class, although the known examples of the first kind are disks.
We leave for future work the construction of supergravity solutions with several
non-commuting monodromies, and the question whether $\Sigma$ is restricted to a disk in that
generality.

The theorems can be restated in terms of the four-dimensional internal space $M_4$,
the $S^2$ fibration over $\Sigma$ that degenerates on $\partial\Sigma$. The fiber is
simply connected, so $b_1(M_4)=b_1(\Sigma)$. For the disk, $M_4$ is a
four-sphere with the points over the five-brane poles removed, and the statement
that $\Sigma$ is a disk is the vanishing of $b_1(M_4)$. After reduction on $\ads_6$,
the Type IIB equations on $M_4$ are those of gravity coupled to a nonlinear sigma
model with target $SL(3,\mathbb R)/SO(2,1)$ and a potential
\cite{Kim:2015AdS6,Kim:2016rhs}. The puncture monodromies act there as symmetries,
with no reference to a preserved charge direction. Whether $b_1(M_4)=0$ follows
from a curvature bound on $M_4$, by a Bochner-type argument, is an open question.

One motivation for the topology question comes from the comparison with
four-dimensional class $\mathcal S$ theories. Fenati, Montero and Valenzuela
\cite[\S4]{Fenati:2026hkn} considered increasing the genus $g_{\mathcal S}$ of the Riemann surface
wrapped by the M5-branes, at fixed large M5-brane number. In the $\ads_5$ dual the internal
space is an $S^4$ fibered over the hyperbolic surface, whose area grows as $g_{\mathcal S}-1$
at fixed curvature radius. The surface itself supplies
no light tower, since the low-lying spectrum of its Laplacian is in general set by the
curvature scale rather than by the area \cite[\S4]{Fenati:2026hkn}. The central charge gives
$(\ell_{\rm AdS}M_{\rm Pl,5})^3\propto g_{\mathcal S}-1$, with $\ell_{\rm AdS}$ the AdS radius and
$M_{\rm Pl,5}$ the five-dimensional Planck mass. The KK tower associated with
the $S^4$ fiber therefore becomes light in five-dimensional Planck units
as $(g_{\mathcal S}-1)^{-1/3}$, while its mass scale in AdS units stays finite. This
supplies the light tower expected from the AdS distance conjecture
\cite{Lust:2019zwm}. In the Type IIB classes
covered by our theorems, the disk restriction excludes this large-genus
limit altogether.

On the disk one can instead increase the number $L$ of boundary poles.
Treating the charges as continuous, consider the rotationally symmetric family
with poles at $p_\ell=e^{2\pi i\ell/L}$ in the unit-disk coordinate $\zeta$
and residues $Z_\ell=q\,e^{-2\pi i\ell/L}$ with $q>0$, so the
charge directions turn clockwise as the poles run counterclockwise.
The stress-tensor coefficient $C_T\propto\int_\Sigma\kappa^2\cG\,d^2\zeta$
\cite[(4.11)]{Gutperle:2018wuk} grows as $L^4$ at fixed $q$, but tends
to a finite limit at fixed $Lq$, the sum of the charge magnitudes. Adding poles at fixed
total charge therefore does not make $C_T$ grow without bound.
For quantized charges, each
external stack contains at least one five-brane, so the total brane count
bounds $L$. In class $\mathcal S$, the genus can grow without bound at
fixed M5-brane count.

At fixed genus, the complex structure moduli of the class $\mathcal S$
surface parametrize exactly marginal couplings of the four-dimensional
theory \cite[\S4]{Fenati:2026hkn}.
By contrast, five-dimensional SCFTs admit no supersymmetry-preserving exactly marginal
deformations \cite{Cordova:2016xhm}. On the disk this is consistent with
the expectation that regularity fixes the cross-ratios of the boundary poles
once the five-brane charges are specified \cite[\S III.1]{Fluder:2020pym}.
An annulus whose modulus was fixed by the charges would be equally
consistent with the absence of a conformal manifold. Our topology theorem
excludes even such an isolated solution in the monodromy-free class.

On an unpunctured disk, a common rescaling of the charges at fixed pole
positions multiplies $\kappa^2$ and $\cG$ by the same factor. The spin-two
spectrum in AdS units is therefore unchanged \cite[(3.12)]{Gutperle:2018wuk}.
As the charges are scaled up, $\ell_{\rm AdS}M_{\rm Pl,6}\to\infty$ and each fixed massive
spin-two KK level becomes light in six-dimensional Planck units.
In AdS units, varying the charges
or the number of poles cannot close the spin-two gap. A mode of $S^2$ angular momentum
$J\ge1$ obeys $m^2\ell_{\rm AdS}^2\ge3J(3J+5)$, with equality for a universal
mode at each $J$ \cite[(3.14), (3.17)]{Gutperle:2018wuk}. For $J=0$ this bound gives only $m^2\ge0$, and the gap comes instead from the $F(4)$ multiplet
classification \cite[Table~22]{Cordova:Multiplets}: a
nonconserved spin-two conformal primary has $\Delta\ge6$, hence
$m^2\ell_{\rm AdS}^2=\Delta(\Delta-5)\ge6$. Thus the lightest massive
spin-two mode on every regular unpunctured disk satisfies
$6\le m_{\rm gap,2}^2\ell_{\rm AdS}^2\le24$, where the upper bound is the universal mode at
$J=1$. The gap is held at the AdS scale. The upper bound excludes conventional scale
separation, in which the gap grows without bound in AdS units, proposed for instance in
\cite{DeWolfe:2005uu}. The lower bound excludes the \emph{inverse scale separation} found, e.g., in
$\ads_4\times S^3\times S^3$ solutions of the non-supersymmetric $O(16)\times O(16)$ heterotic
string. There, for unequal fluxes, one $S^3$ becomes parametrically larger than the
AdS radius \cite{Basile:2026Inverse}, so that the gap set by its radius closes in AdS units.

In the $\ads_5$ duals of class $\mathcal S$ theories, on the other hand, the spin-two gap
is not held at the AdS scale. The difference does not come from the supergravity bounds.
These duals are solutions of eleven-dimensional supergravity in which the
$SU(2)_R\times U(1)_r$ R-symmetry acts as the isometry of an $S^2$ and a circle, and Chen,
Gutperle and Uhlemann found a lower bound on the spin-two masses that is fixed by the two
R-charges and saturated by universal modes \cite[(2.40)]{Chen:2019ydk}. For R-symmetry
singlets it reduces to $m^2\ge0$, just as the $\ads_6$ bound $3J(3J+5)$ does at $J=0$.
In neither case does the supergravity bound say how light a massive singlet can be. That is
decided by the superconformal algebra and by the internal geometry.

Consider the simplest class $\mathcal S$ case, M5-branes wrapped on a compact hyperbolic surface.
The spin-two masses follow from a separation of variables
\cite[(22)]{Bhattacharya:2024tjw}. Modes that are constant on the $S^4$ fiber are
R-symmetry singlets and have
$m^2\ell_{\rm AdS}^2=2\lambda_{\mathcal S}$, with $\lambda_{\mathcal S}$ an
eigenvalue of the Laplacian of the surface in the metric of curvature $-1$. The lightest
massive mode of this kind is set by the first nonzero eigenvalue. As noted above, this
eigenvalue is in general set by the curvature scale. It depends on the complex
structure, however, and becomes parametrically small when closed geodesics that separate
the surface are pinched \cite{Schoen:1980}.
In the four-dimensional theory, this is the limit in which the theory separates into
decoupled sectors, each with its own stress tensor. The four-dimensional $\mathcal N=2$
algebra allows the gap to close, since a nonconserved spin-two operator only needs
$\Delta>4$, that is $m^2>0$, whereas the $F(4)$ algebra requires $\Delta\ge6$
\cite{Cordova:Multiplets}.

Taken together, these comparisons point to a rigidity of the $\ads_6$ solutions dual to
five-dimensional SCFTs, in contrast with the $\ads_5$ duals of class $\mathcal S$ theories.
There the genus of the surface can grow
without bound, and its complex structure spans a conformal manifold along which
the spin-two gap can close. In the
monodromy-free $\ads_6$ class, regularity fixes the topology, the dual theories
admit no exactly marginal deformations, and the $F(4)$ superconformal symmetry holds the
spin-two gap at the AdS scale. The external five-brane charges are then expected
to be the only data left to vary. It would be interesting to learn whether smeared
sources relax any part of this rigidity.

\medskip
\noindent\textbf{Acknowledgments:} The author thanks Hyojoong Kim and Minwoo Suh for discussions and for an earlier collaboration on $\ads_6$ solutions of Type IIB supergravity, and Se-Jin Kim for suggesting to use Lean to check the proofs. This work is supported by the National Research Foundation of Korea under grants RS-2025-00518906, RS-2025-25441170, and RS-2025-25457100.

\medskip
\noindent\textbf{AI statement:} The author used Claude Code (Anthropic) and
Codex (OpenAI) to help check derivations, search the literature, prepare figures and revise the manuscript. Parts of the argument were formalized in Lean~4 with Mathlib, with the code written by
Claude Code: the counting steps of \cref{thm:onlydisk,thm:sevenbranes,thm:extended}, the pointwise
identities, the non-vanishing of $\lambda'$ at boundary points, and the rotation bound of
\cref{lem:rotation}. The remaining analytic inputs, including Hopf's lemma for $\cG$ and the degree
formula on the double, enter there as hypotheses. The Lean files are included as ancillary files.
The author verified the results and proofs and takes responsibility
for the final text.

\IfFileExists{JHEP.bst}{\bibliographystyle{JHEP}}{\bibliographystyle{unsrt}}
\bibliography{refs}

\end{document}